\documentclass{aa}

\usepackage{svg}
\usepackage{graphicx}
\usepackage{siunitx}
\usepackage{comment}						 
\includecomment{comment}
\specialcomment{note}{\begingroup\sffamily\color{gray}\footnotesize}{\endgroup}

\usepackage{txfonts}
\usepackage{color}
\definecolor{snsgreen}{rgb}{0.0, 0.620, 0.451}
\newcommand{\rev}[1]{{#1}}
\newcommand{\revii}[1]{{#1}}

\excludecomment{note}                       

\usepackage[disable]{todonotes}			

\usepackage[colorlinks]{hyperref}
\hypersetup{colorlinks=true,linkcolor=blue,citecolor=blue,filecolor=blue} %

\DeclareSIUnit\mSun{M_\odot}
\DeclareSIUnit\Msun{M_\odot}
\DeclareSIUnit\mStar{M_\star}
\DeclareSIUnit\Mstar{M_\star}
\DeclareSIUnit\mEarth{M_\oplus}
\DeclareSIUnit\Mearth{M_\oplus}
\DeclareSIUnit\rEarth{R_\oplus}
\DeclareSIUnit\Rearth{R_\oplus}
\DeclareSIUnit\year{yr}
\DeclareSIUnit\au{au}
\DeclareSIUnit\dex{dex}
\DeclareSIUnit\ppm{ppm}
\DeclareSIUnit\eV{eV}
\newcommand{\msini}{M \sin (i)}

\begin{document}

\newcommand{\Nstarsharps}{\ensuremath{77}}
\newcommand{\Nstarscarmenes}{\ensuremath{70}}
\newcommand{\Nstars}{\ensuremath{147}}
\newcommand{\Nduplicatetargets}{\ensuremath{22}}
\newcommand{\Nplanetsharps}{\ensuremath{8}}
\newcommand{\Nplanetscarmenes}{\ensuremath{27}}
\newcommand{\Nplanets}{\ensuremath{35}}
\newcommand{\Nplanetssystems}{\ensuremath{26}}
\newcommand{\multPlanets}{\ensuremath{1.35}}
\newcommand{\NplanetsNGMzero}{\mbox{2,433,170}}
\newcommand{\Ndetectable}{\mbox{20,021}}
\newcommand{\Ndetectablesystems}{\mbox{14,513}}
\newcommand{\multDetectable}{\mbox{1.38}}
\newcommand{\ntotobs}{\ensuremath{0.24}}
\newcommand{\ntotobserr}{\ensuremath{0.04}}
\newcommand{\ntotsyn}{\ensuremath{0.201}}
\newcommand{\ntotsynerr}{\ensuremath{0.001}}
\newcommand{\Ngiants}{\ensuremath{4}}
\newcommand{\adstatisticMsini}{\ensuremath{0.97}}
\newcommand{\adstatisticPeriods}{\ensuremath{7.85}}
\newcommand{\Mthresh}{\ensuremath{0.4}}
\newcommand{\dipPvalNGM}{\ensuremath{\SI{7e-3}{}}}
\newcommand{\dipPvalNGMEarly}{\ensuremath{\SI{1e-3}{}}}
\newcommand{\dipPvalNGMLate}{\ensuremath{0.9}}
\newcommand{\dipPvalobs}{\ensuremath{0.8}}

    \title{RV-detected planets around M~dwarfs: Challenges for core accretion \revii{models}}

\author{M. Schlecker\inst{\ref{inst:mpia},\ref{inst:ua}}
  \and R. Burn\inst{\ref{inst:unibe},\ref{inst:mpia}}
  \and S. Sabotta\inst{\ref{inst:lsw}}
  \and A. Seifert\inst{\ref{inst:mpia}}
  \and Th. Henning\inst{\ref{inst:mpia}}
  \and A. Emsenhuber\inst{\ref{inst:usm},\ref{inst:ualpl},\ref{inst:unibe}}
  \and C. Mordasini\inst{\ref{inst:unibe}}
  \and S. Reffert\inst{\ref{inst:lsw}}
  \and Y. Shan\inst{\ref{inst:oslo},\ref{inst:goe}}
  \and H. Klahr\inst{\ref{inst:mpia}}
          }

\institute{Max-Planck-Institut für Astronomie, Königstuhl 17, 69117 Heidelberg, Germany\\
\email{schlecker@arizona.edu} \label{inst:mpia}
\and
\rev{Department of Astronomy/Steward Observatory, The University of Arizona, 933 North Cherry Avenue, Tucson, AZ 85721, USA} \label{inst:ua}
\and
Physikalisches Institut, University of Bern, Gesellschaftsstrasse 6, 3012 Bern, Switzerland \label{inst:unibe}
\and
Landessternwarte, Zentrum für Astronomie der Universit\"at Heidelberg, K\"onigstuhl 12, 69117 Heidelberg, Germany \label{inst:lsw}
\and
Lunar and Planetary Laboratory, University of Arizona, 1629 E. University Blvd., Tucson, AZ 85721, USA \label{inst:ualpl}
\and
Universitäts-Sternwarte München, Ludwig-Maximilians-Universität München, Scheinerstraße 1, 81679 München, Germany \label{inst:usm}
\and
\rev{Centre for Earth Evolution and Dynamics, Department of Geosciences, University of Oslo, Sem Sælands vei 2b 0315 Oslo, Norway} \label{inst:oslo}
\and
Institut für Astrophysik, Georg-August-Universität, Friedrich-Hund-Platz 1, 37077 Göttingen, Germany \label{inst:goe}
}

   \date{\today}

 
  \abstract
   {Planet formation is sensitive to the conditions in protoplanetary disks, for which scaling laws as a function of stellar mass are known.}
   {We aim to test whether the observed population of planets around low-mass stars can be explained by these trends, or if separate formation channels are needed.}
   {We address this question by confronting \rev{a state-of-the-art planet population synthesis model} with a sample of planets around M~dwarfs observed by the HARPS and CARMENES radial velocity (RV) surveys.
   To account for detection biases, we performed injection and retrieval experiments on the actual RV data to produce synthetic observations of planets that we simulated following the core accretion paradigm.}
   {These simulations robustly \rev{yield the previously reported} high occurrence of rocky planets around M~dwarfs and generally agree with their planetary mass function.
   In contrast, \rev{our simulations cannot reproduce a population of giant planets around stars less massive than 0.5 solar masses.}
   This \rev{potentially indicates an} alternative formation channel for giant planets around the least massive stars that cannot be explained with current core accretion theories.
   We further find a stellar mass dependency in the detection rate of short-period planets. A lack of close-in planets around \rev{the earlier-type} stars ($M_\star \gtrsim \SI{0.4}{\Msun}$) \rev{in our sample} remains unexplained by our model and \rev{indicates} dissimilar planet migration barriers in disks of different spectral subtypes.}
   {Both discrepancies can be attributed to gaps in our understanding of planet migration in nascent M~dwarf systems.
   They underline the different conditions around young stars of different spectral subtypes, \rev{and the importance of taking these differences into account when studying} planet formation.}


   \maketitle
%
\section{Introduction}
M~dwarf stars are not only the most abundant stars and planet hosts in the solar neighborhood~\citep{Reyle2021, Hsu2020}, but also rewarding study objects regarding some of the most pressing questions on planet formation.
They have been shown to host many small, potentially habitable planets~\citep[e.g.,][]{Ribas2018,Bluhm2020,Ment2020,Kemmer2020,Cloutier2021,Kossakowski2021}, and much has been learned about their properties from recent demographic studies~\citep[e.g.,][]{Mulders2015a,Dressing2015,Gibbs2020,Hsu2020}.
In particular, the occurrence of small planets appears to increase with decreasing stellar mass \citep{Mulders2015,Hardegree-Ullman2019}, and they might orbit closer to their hosts at very low stellar masses~\citep{Sabotta2021}.
On the other hand, the frequency of giant planets around M~dwarfs seems to be low\rev{~(\citealp[][but see]{Endl2006,Johnson2010,Bonfils2013,Obermeier2016,Ghezzi2018} \citealp{Jordan2022})}.
To understand how such trends come about, it is crucial to confront observational results with predictions made by planet formation theory.
In this paper, we perform such a comparison using a well-defined observational sample and thorough characterization of its biases, as well as a planet population synthesis tailored to that sample.

On the observational side, the tightest constraints on exoplanet demographics have been based on data from the \emph{Kepler} transit survey \citep{Borucki2010}, which detects planets down to very low masses and radii.
However, \emph{Kepler} has observed only a few thousand stars of spectral type~M \citep[e.g.,][]{Gaidos2016}, and transit observations do not provide information about the mass of a planet.
\rev{Instead of planetary radii, which are degenerate for giant planets~\citep[e.g.,][]{Hatzes2015}, radial velocity (RV) surveys constrain projected planet masses.
For comparisons to planet formation models, this is the more fundamental quantity due to its direct relationship to planetary accretion processes.}
RV surveys \rev{also} require less restrictive geometrical configurations of detectable planetary systems, enabling more complete observations.
They thus cover a unique parameter space and provide complementary input to exoplanet demographics.
Spectroscopic surveys are able to identify \rev{active stars~\citep[e.g.,][]{TalOr2018,Jeffers2018} and binary stars~\citep[e.g.,][]{Baroch2018,Baroch2021}}, whose implicit or explicit exclusion from catalogs affect the planet statistics~\citep{Moe2021}.
This makes this detection technique ideal for the comparison with current population synthesis models, which \rev{currently} do not include stellar multiplicity or activity.

One of the most comprehensive \rev{datasets} currently available have been generated by the High Accuracy Radial velocity Planet Searcher~\citep[HARPS,][]{Mayor2003} and the Calar Alto high-Resolution search for M~dwarfs with Exoearths with Near-infrared and optical Echelle Spectrographs~\citep[CARMENES,][]{Quirrenbach2010,Quirrenbach2013,Reiners2018} surveys.
\cite{Bonfils2013} reported planet occurrence rates for 102~low-mass stars and 14 planets, for which they took approximately 2000 spectra with the HARPS instrument.
A more recent study by~\citet{Sabotta2021} determined occurrence rates using $\sim$6500 \rev{CARMENES} spectra of 71 M~dwarfs that host 27 planets.
Combining these two surveys results in 9 confirmed planets around \Nstars~target stars in the regime of \SIrange{0.5}{2.5}{\Rearth} and \SIrange{0.5}{10}{\day} where \emph{Kepler} found 13 confirmed planets around 561 target stars~\citep{Hardegree-Ullman2019}, thus the overall detection rate is higher for the RV sample.

From the perspective of formation theory, the first fully operational planet population syntheses focused on isolated gas giant planets around solar-type stars~\citep{Ida2005,Alibert2011}.
As theoretical understanding, computational resources, and algorithms improved, more and more physical mechanisms where added to the models, among them planet migration~\citep{Ida2008,Dittkrist2014}, long-term evolution of planet interiors~\citep{Mordasini2012}, atmospheric escape~\citep{Jin2014}, \rev{pebble accretion~\citep{Brugger2018,Bitsch2019a},} and disk chemistry~\citep{Cridland2016,Thiabaud2014}.
A major advancement was to simulate in the same disk multiple planets that gravitationally interact~\citep{Thommes2008,Ida2010,Alibert2013}, which is particularly important for realistic modeling of systems including small planets~\citep{Mordasini2018}.
Models including such an \mbox{N-body} component are capable of computing systems that include the whole range of planetary masses.
Most population synthesis efforts have limited themselves to planetary systems around stars with solar mass, but some studies took into account stellar mass dependencies~\citep{Ida2005,Alibert2011,Liu2019,Miguel2020,Burn2021}.
The \rev{planetesimal accretion-based} formation model of~\citet{Emsenhuber2021} used here includes \rev{all above mechanisms except for pebble accretion}, and its extension to low-mass stars presented in \citet{Burn2021} allows to cover the full M~dwarf mass range.

Direct comparison of its synthetic planets with observed samples is impeded by various selection effects and detection biases, \rev{some of which} can be corrected.
A common approach is to ``de-bias'' the observed sample, for example by weighting individual planet discoveries according to their detection probability \citep[e.g.,][]{Cumming2008} or by injecting a test planet population to obtain the necessary correction factors \citep[e.g.,][]{Bonfils2013}.
The drawback of such techniques is that slightly different assumptions can affect the calculated result, typically an occurrence rate in planetary parameter space~\citep{Hsu2020, Sabotta2021}.
A method that does not require extrapolations beyond the actually observed sample is to apply a detection bias to the synthetic planet population~\citep[e.g.,][]{Mulders2019}, hence this is our method of choice.

In this paper, we compile a combined planet sample from the HARPS~\citep{Mayor2003} and CARMENES~\citep{Quirrenbach2010,Quirrenbach2013,Reiners2018} M~dwarf surveys.
Instead of applying a simple RV \rev{amplitude} cut, we use the actual RV measurements for each star to compute the detection bias and apply it to our synthetic population to simulate its observation.
We further take into account the stellar mass distribution of the combined survey.
Finally, we confront the observed and biased synthetic planet populations in minimum mass-period-stellar mass-space.
\rev{The main purpose of our comparison of detection rates, planetary masses, and orbital period distributions is to show which of the CARMENES and HARPS planet detections our model can and cannot explain.}

The paper is organized as follows:
First, we introduce the observed samples in Sect.~\ref{sec:observed_sample}, followed by a brief description of the formation model and synthetic planet population in Sect.~\ref{sec:synthetic_sample}.
We continue with presenting statistical comparisons of the observed and simulated samples in Sect.~\ref{sec:results}, and discuss the most relevant differences in \ref{sec:discussion}.
In Sect.~\ref{sec:conclusions}, we provide conclusions to our findings.


\section{Observed sample}\label{sec:observed_sample}
   We compiled a sample of confirmed exoplanets orbiting M~dwarf stars.
   By combining results from two of the largest RV surveys targeting low-mass stars, the HARPS survey~\citep{Bonfils2013} and the CARMENES search for exoplanets around M~dwarfs~\citep{Quirrenbach2010}, we obtained an observed sample of \Nplanets\ planets around \Nstars\ stars.
In the following, we describe how we compiled this sample.

\subsection{The CARM$_{70}$ sample}
One of the most comprehensive searches for exoplanets around M~dwarfs is the CARMENES high-precision RV survey.
The CARMENES \textit{instrument} consists of two independent Echelle spectrographs, one for visual wavelengths \SIrange{0.55}{1.05}{\micro\meter} and one for near-Infrared wavelengths \SIrange{0.95}{1.7}{\micro\meter}~\citep{Quirrenbach2013}.
Both channels are fiber-fed from the Calar Alto \mbox{3.5~m telescope}.
In its Guaranteed Time Observations (GTO), CARMENES targets a sample of $\sim350$ stars whose spectral type distribution peaks at M4V~\citep{Reiners2018}.
This survey started beginning of 2016 and has since produced more than 18,000 spectra~\citep{Sabotta2021} and led to various exoplanet discoveries~\citep[e.g.,][]{Sarkis2018,Ribas2018,Luque2018,Morales2019,Zechmeister2019,Stock2020,Nowak2020,Trifonov2021}.
The survey has already been defined with the goal to perform a population-level, statistical analysis on the datasets it produces.
For a subset of the GTO stars, at least 50~RV measurements \rev{have been collected} and observations are considered complete~\citep{Sabotta2021}.
In addition to the constraint on the RV measurements, all spectroscopic binaries \citep{Baroch2018} were excluded as well as a set of very active stars with large RV scatter \citep{TalOr2018}.
After accounting for duplicate stars in our combined stellar sample (see Sect.~\ref{sec:combined_starsample}), a subset of 70 stars, here termed CARM$_{70}$, remains.
Several of the CARMENES discoveries were published in combination with other instruments.
\rev{To ensure a homogeneous analysis, \citet{Sabotta2021} thus used only CARMENES data in their automated search for planetary signals with a false alarm probability (FAP) of less than 1\,\% in the Generalized Lomb-Scargle periodograms (GLS) \citep{Zechmeister2009}.}
\rev{There, we used a conservative period cutoff at half the time baseline to make sure that at least two orbits of the planet were observed.
In this way, we identified 27 planets in 22 planetary systems.}

\subsection{The HARPS M~dwarf sample}
The High Accuracy Radial velocity Planet Searcher (HARPS) consists of an Echelle spectrograph fed from the ESO La Silla \mbox{3.6~m telescope} with a second fiber for a reference spectrum \citep{Mayor2003}. It observes stars in different mass regimes, reaching from \SI{0.1}{\Msun}~\citep{Bonfils2013} up to intermediate masses \citep{Leao2018}.
The original M~dwarf subsample of the HARPS search for exoplanets \rev{comprises a volume-limited ($< 11\,\mathrm{pc}$) target list, which was further cropped by declination ($\delta < +20\deg$) and magnitude ($V > 14\,\mathrm{mag}$).
It} contains 102 stars covering a mass range of mostly \SIrange{0.1}{0.6}{\Msun}~\citep{Bonfils2013}.
Their stellar masses were calculated from empirical mass-luminosity relations~\citep{Delfosse2000}.
As in \cite{Bonfils2013}, we excluded four stars with less than four RV measurements from the sample and additionally one target with five very low signal-to-noise observations (L~707-74).
A scarcity of stars between \SI{\sim 0.35}{\Msun} and \SI{0.40}{\Msun} in their sample has been shown to be a statistical fluctuation~\citep{Neves2013}.
Unlike CARM$_{70}$, the HARPS M~dwarf sample was not filtered to exclude very active stars.
For this reason the stellar mass distribution of the HARPS sample has a lower median stellar mass of \SI{0.29}{M_\odot}.
In the planet sample of \cite{Bonfils2013} there were 14 planets. Several of those planets were published \rev{as a result of analyzing a} combination of time series with other instruments. Therefore, for our analysis, we retrieved again those signals that were present in HARPS data only. After this analysis, we excluded the two planets \object{Gl 849b} and \object{Gl 832b} from the sample, because the time baseline of the time series did not cover two whole orbits (they were published in combination with data from other instruments). \object{GJ 876d} was also excluded from our analysis as it is below the detection limit of HARPS data alone. Furthermore, we excluded \object{GJ 581d} as it is probably a false positive \citep{Robertson2014,Hatzes2016}.

\subsection{The combined M~dwarf sample}\label{sec:combined_starsample}
\Nduplicatetargets\ stars occur in both the HARPS and the CARMENES target list.
To avoid duplicates while preserving sensitivity, we determined for each star the number of observations it received for each instrument and kept \rev{in the sample the dataset} with the greater number of RV values.
Some of these duplicate stars host planets, in which case we considered only the planets detectable \rev{with the retained dataset.}
No planets were lost from the sample due to this rule.
In total, the combined HARPS\&CARM$_{70}$ sample consists of \Nstars\ target stars and \Nplanets\ planets.
All targets and duplicate stars in our sample are listed in Table~\ref{table:starlist}.
Not only the mass distribution of the stellar sample is relevant for demographic assessments, but also how intensely each target was monitored.
We show both the mass distributions of each stellar subsample and the number of observations for each mass bin in Fig.~\ref{fig:Mstar_hist}\footnote{For the distribution of the numbers of spectra in spectral subtype we refer to Fig.~10 in~\citet{Sabotta2021}.}.
Some of the planets have new parameter constraints from follow-up studies, in which case we used the updated values (see Table~\ref{table:planets1mchapter}).

\section{Synthetic planet population}\label{sec:synthetic_sample}
We computed a synthetic planet population using the Bern global model of planetary formation and evolution.
Here, we briefly summarize the methodology, which has been described extensively in the previous works by \citet{Emsenhuber2021}, \citet{Emsenhuber2021b}, \citet{Schlecker2021}, \citet{Burn2021}, \citet{Schlecker2021b}, and \citet{Mishra2021}.

\subsection{Formation and evolution model}
The goal of a global model is to include all relevant physical processes that occur in the formation and evolution, albeit in a simplified fashion, so that their interactions can be studied.
The current version of the model is based on several previous generations \citep{Alibert2005,Mordasini2009,Alibert2013}.
Its formation part computes concurrently the evolution of a protoplanetary disk consisting of gas and planetesimals, the accretion of solids and gas by the protoplanets, gas-driven migration, and the dynamical interactions between multiple protoplanets.
The starting point of the simulations is an already-formed disk consisting of a gaseous part with mass $M_{\rm disk}$, a dust component (1\%) used to calculate the opacity, and a disk of planetesimals with mass $M_{\rm plts}$.
The gas disk is modeled as an axissymmetric 1D radial profile following \citet{Lust1952} and \citet{Lynden-Bell1974}, where the vertical structure is obtained following the procedure of \citet{NakamotoNakagawa1994} with direct irradiation from the star on the surface \citep{Hueso2005}.
Turbulent viscosity is described by an $\alpha$ parametrization \citep{Shakura1973}.
In addition, both internal \citep{Clarke2001} and external \citep{Matsuyama2003} photoevaporation are included.
The combination of stellar accretion and photoevaporation leads to the dispersal of the gas disk after a few~\SI{}{\mega\year}, comparable to observed disk lifetimes~\citep[e.g.,][]{Haisch2001,Mamajek2009,Ribas2015}.

Planets are assumed to form according to the core accretion paradigm \citep{Perri1974,Mizuno1980}.
At the start of each simulation, we inserted multiple planetary seeds with an initial mass of \SI{0.01}{M_{\oplus}}.
Their cores grow by accretion of planetesimals in the oligarchic regime \citep{Ida1993,Inaba2001,Fortier2013}.
Gas accretion rates are determined by the ability of the planet to cool by radiating away energy and, for planets exceeding a critical mass, by the gas supply from the disk.
We computed nominal accretion rates by solving the internal structure equations of both the solid planetary core and the gaseous envelope \cite{Bodenheimer1986}, where we assumed an onion-like structure, the deposition of accreted solids at the core-envelope boundary, and spherical symmetry.
The maximum accretion rate that can be provided by the gas disk is given by the Bondi accretion in two or three dimensions, that is, the amount of gas that is intersected by the planet with a gas capture radius larger (two-dimensional problem) or smaller (three-dimensional problem) than the local disk scale height~\citet{Dangelo2008}.
Protoplanets embedded in a gas disk will undergo planetary migration.
We account for \mbox{type-I} migration following \citealp{Paardekooper2011} with a modulation due the planet's orbital characteristics from \citealp{ColemanNelson2014}.
For very massive planets, we model \mbox{type-II} migration following \citet{Dittkrist2014}.
Planet eccentricities and inclinations are damped by the gas disk, following the prescription of \citet{CresswellNelson2008} in \mbox{type-I} migration and a fraction of the migration timescale in the \mbox{type-II} regime.
Dynamical interactions between the protoplanets are tracked by means of the \texttt{mercury} \textit{N}-body package \citep{Chambers1999}, which includes a prescription of collisions described in \citet{Emsenhuber2021}.
Beyond the gas disk phase, planetesimals accretion and orbital evolution are tracked until the end of the formation stage at \SI{20}{\mega\year}.
In the following evolution stage the thermodynamical evolution is tracked until \SI{10}{\giga\year}~\citep{Mordasini2012} .
This stage includes migration due to stellar tides for close-in planets \citep{Benitez-Llambay2011} and photoevaporation to remove atmospheric material \citep{Jin2014}.
\rev{While comprehensive, the Bern model currently lacks prescriptions for pebble accretion and the formation of planetesimals and planet seeds, both of which are currently in development~\citep{Brugger2020,Voelkel2020b,Voelkel2021}}.

\subsection{The synthetic \rev{M~dwarf} planet population}\label{sec:BernModel_mdwarf}
The population synthesis approach randomizes initial conditions to sample the parameter space.
In a Monte Carlo fashion, $M_{\rm disk}$ and $M_{\rm plts}$, as well as the external photoevaporation rate, the inner disk edge, and the starting location of each seed were varied between each run of the model.
For the M~dwarf planet population, termed \textit{NGM}, we chose discrete stellar masses of 0.1, 0.3, 0.5, and \SI{0.7}{M_{\odot}}~\citep{Burn2021}.
\rev{Some of the disk parameters have been shown to scale with stellar mass, and we aimed to scale our Monte Carlo parameter distributions accordingly.
For the disk mass, we} extrapolated the observed relation between disk mass and stellar mass found by \citet{Pascucci2016,Ansdell2017} to earlier times and adopted a linear relationship between star and disk masses.
\rev{This choice of scaling is supported by observed stellar accretion rates \citep{Alcala2017}, which compare reasonably well to the model output from \citet[][see their Fig. 3]{Burn2021}.
In order to account for the observed scatter in the measured disk masses of nearby star forming regions, we distributed our gas disk masses closely following the reported masses of Class-I objects in~\citet{Tychoniec2018}, resulting in a log-normal distribution $\mathcal{N}(\mu=-1.49,\,\sigma^2=0.123)$~\citep[compare][]{Emsenhuber2021b,Schlecker2021}.
}
The total planetesimal mass $M_{\rm plts}$ directly relates the gas disk mass $M_{\rm disk}$ with a dust-to-gas ratio, which in turn translates to a stellar metallicity under the assumption that stellar and disk metallicity are the same~\citep{Emsenhuber2021b}.
We \rev{distributed} this parameter normally according to [Fe/H] measurements in the solar neighborhood~\citep{Santos2003}.
\rev{We assumed that inner disk edges are} located at the co-rotation radius with the stellar spin \citep{Guenther2013}.
In agreement \rev{with measured T~Tauri rotation rates~\citep{Venuti2017}, our inner edges are distributed log-normally ($\mu = \SI{4.74}{\day}, \sigma^2 = 0.3\,\mathrm{dex}$) in orbital period and do not scale with stellar mass.}
At each simulation run, we injected 50~planetary seeds into the disk at random radial positions between the inner edge and an orbital period of \SI{253}{\year}.
The external photoevaportion rate of the gas disk is chosen such that the resulting lifetimes of the disks are similar for all stellar masses with a median $\sim$\SI{3.0}{Myr}.
More details on the distributions of the Monte Carlo parameters can be found in \citet{Burn2021}.

Other parameters are fixed for each simulation run:
the viscous $\alpha$ parameter was set to \SI{2e-3}{}, and the initial slope of the planetesimal and gas disk are -1.5 and -0.9, respectively.
This results in a high concentration of solid mass close to the star, where locally, planetesimals can make up to \SI{10}{\percent} of the total disk mass.
Furthermore, we specified a planetesimal diameter of \SI{600}{\meter}.
This relatively low value leads to efficient damping by gas drag and therefore lower relative velocities between the growing planet and the planetesimals.
A more detailed discussion on planetesimal sizes can be found in \citet{Emsenhuber2021b}.
In total, the \textit{NGM} population consists of 4996 simulated systems. 
\begin{table*}[htbp]
    \caption[]{Synthetic planet population \textit{NGM}. Each of the simulated systems started with \mbox{$N_\mathrm{emb,ini}=50$} planetary seeds. Five stellar masses and corresponding effective temperatures were sampled with different weights to match the observed distribution, and a total of 100,000 systems were drawn. Each system was assigned a random, isotropic inclination angle $i$. The population NG75 extends \textit{NGM} to solar-type stars~\citep[see][]{Emsenhuber2021b,Burn2021} but did not have to be sampled to match the $M_\star$ distribution of the observed sample. Table adapted from~\citet{Burn2021}.}
    \label{tab:NGM_masses}
    \centering
    \begin{tabular}{c c c c c c c}
        \hline\noalign{\smallskip}
        Name & $M_\star$ & $T_{\mathrm{eff}\mathrm{, 5 Gyr}}^{(a)}$  & $N_\mathrm{emb,ini}$ & \multicolumn{1}{c}{\centering Simulated Systems} & \multicolumn{1}{p{2cm}}{\centering Sampling Weights} & \multicolumn{1}{p{2cm}}{\centering Resampled Systems}\\
        \hline\hline\noalign{\smallskip}
        NGM10 & \SI{0.1}{\mSun} & \SI{2811}{\kelvin} & 50 & 1000 & 0.204 & 21392\\
        NGM14 & \SI{0.3}{\mSun} & \SI{3416}{\kelvin} & 50 & 997 & 0.368 & 34768\\
        NGM11 & \SI{0.5}{\mSun} & \SI{3682}{\kelvin} & 50 & 1000 & 0.388 & 39862\\
        NGM12 & \SI{0.7}{\mSun} & \SI{4430}{\kelvin} & 50 & 999 & 0.039 & 3978\\
        NG75$^{(b)}$  & \SI{1.0}{\mSun} & \SI{5731}{\kelvin} & 50 & 1000 & 0 & 0\\
        \hline
        \multicolumn{5}{l}{$^{(a)}$ following \citet{Baraffe2015}} \\
        \multicolumn{5}{l}{$^{(b)}$ population also discussed in~\citet{Emsenhuber2021b}.} \\
        \hline
    \end{tabular}
\end{table*}
Table~\ref{tab:NGM_masses} lists the simulation runs for each host star mass with the corresponding stellar effective temperature at a simulation time of \SI{5}{\giga\year}, $T_{\mathrm{eff}\mathrm{, 5 Gyr}}$.
The initial number of planetary seeds per system $N_\mathrm{emb,ini}$ was always 50.

\subsection{Observing the synthetic population} 
In order to make a comparison to observed data meaningful, we had to take into account their selection effects and detection biases.
We implemented this by weighted resampling of the simulated population and by performing injection-and-retrieval experiments with the observed data.

\subsubsection{Stellar mass sampling}
\begin{figure}
    \centering
    \includegraphics[width=\hsize]{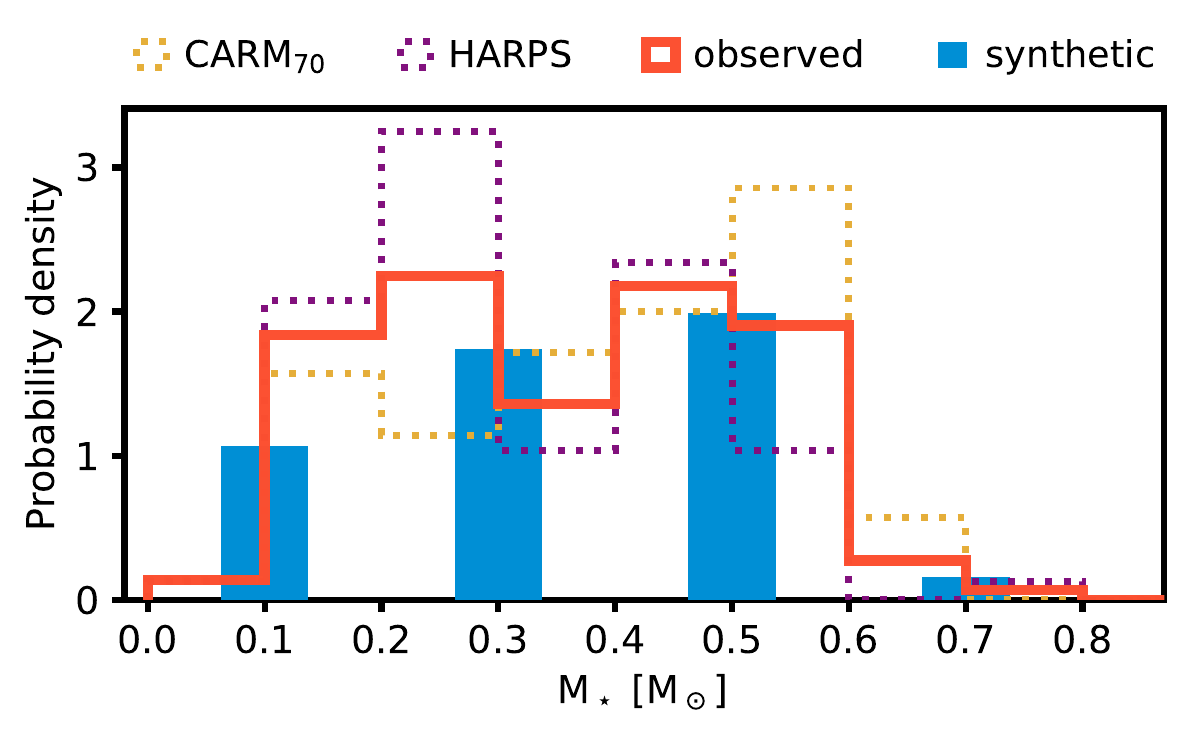}
    \includegraphics[width=.95\hsize]{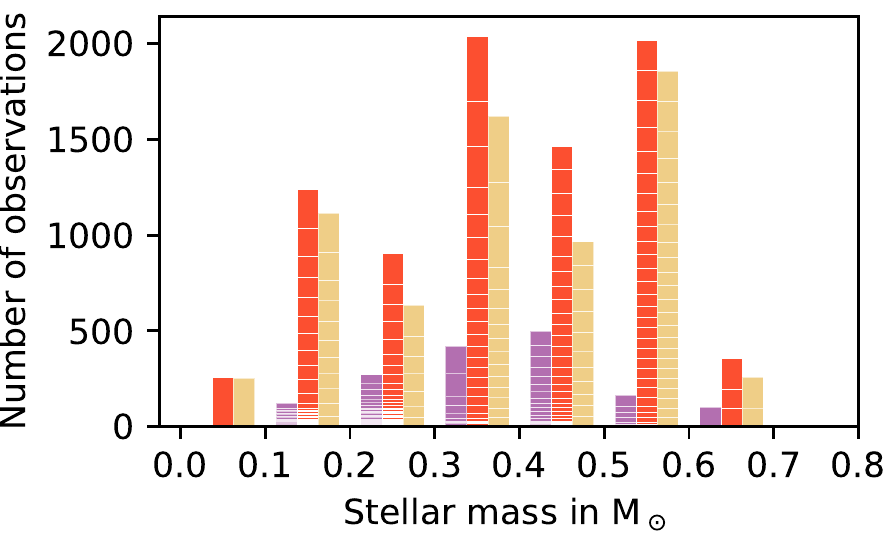}
    \caption[]{Observed and simulated stellar sample. Top: The combined observed stellar mass distribution is composed of different distributions for the HARPS (purple) and CARM$_{70}$ samples (yellow), respectively. The host star masses in the synthetic \textit{NGM} population are discrete\rev{.} By weighted resampling of its systems according to the combined observed sample (red), we obtain a new distribution that approximates it (blue). Bottom: Number of observations per stellar mass bin with the same color code as the top figure. Every target star is represented by a white box.}
    \label{fig:Mstar_hist}
\end{figure}
In contrast to the \textit{NGM} population, the HARPS\&CARM$_{70}$ sample has a continuous distribution of stellar masses, as shown in Fig.~\ref{fig:Mstar_hist}.
We approximated this distribution by weighted resampling of our synthetic population.
For this purpose, we first computed a histogram of the observed stellar mass sample with the bin edges defined as the center between the discrete \textit{NGM} stellar masses.
The normalized histogram counts then served as sampling weights, which for HARPS\&CARM$_{70}$ amount to 0.204, 0.368, 0.388, 0.039, and 0 for \textit{NGM}'s host star masses \SIrange{0.1}{1.0}{\Msun} (see Tab.~\ref{tab:NGM_masses}); that is, the \SI{1.0}{\Msun} population has no contribution.
In total, we sampled 100,000 systems with replacement.
As the synthetic planets will be compared to RV-detected exoplanets, for which only minimum masses $\msini$ are known, we assigned them random orbital inclination angles $i$.
Here, we assumed an isotropic distribution of orbit orientations and, for each system, drew a $\sin(i)$ term from the distribution~\citep{Ho2011}
\begin{equation}
        f(\sin(i)) = \frac{\sin(i)}{\sqrt{1 - \sin(i)^2}}.
        \label{eq:random_sini}
\end{equation}
Hence, despite our oversampling of the \textit{NGM} population, no planet occurs more than once with the exact same properties.
The resulting $M_\star$ distribution approximates the one of the HARPS\&CARM$_{70}$ sample (compare Fig.~\ref{fig:Mstar_hist}).
While the oversampled population contains \SI{5e6}{} planets, in the following we consider only the \NplanetsNGMzero\ planets that survived the formation and evolution phase until an assumed observation time $t_\mathrm{obs}=\SI{5}{\giga\year}$.

\subsubsection{Accounting for detection bias}
Exoplanet surveys are affected by detection biases, distorting the distributions in planetary parameter space.
Some of these biases can be characterized and corrected, which is commonly done for occurrence rate studies~\citep[e.g.,][]{Cumming2008,Zechmeister2009,Mayor2011,Bonfils2013,Sabotta2021, Fulton2021}.
To obtain a synthetic planet population that we can reasonably compare to our observed sample, we applied the same bias to the \textit{NGM} planets.
In this way, we obtain an observable \textit{NGM} population that contains all the planets that a combined HARPS\&CARM$_{70}$ survey would have detected if it observed our synthetic systems.

\rev{In~\citet{Sabotta2021}, we} conducted injection-and-retrieval experiments on the RV data for each of the stars in the CARM$_{70}$ sample.
The period and $\msini$ of the injected planets were taken from a grid of log-uniform distribution.
For every grid point\rev{, we} injected 50 test planets with random phase angles.
If the test planet appeared in the GLS periodogram with a FAP of less than 1\,\%, \rev{we counted it} as a recovery.
The detection probability for this grid point is then the number of recovered planets divided by the total number of injected planets. From the final survey sensitivity map, we \rev{excluded active periods that show up in the periodogram of the activity indicators.
We further dismissed} periods that are longer than half the time baseline by setting the detection probability to zero. At those periods we would not accept a real \rev{planetary} signal (see Sect.~\ref{sec:observed_sample}), and therefore we do not accept \rev{corresponding} test planets.

\begin{figure*}
    \centering
    \includegraphics[width=\hsize]{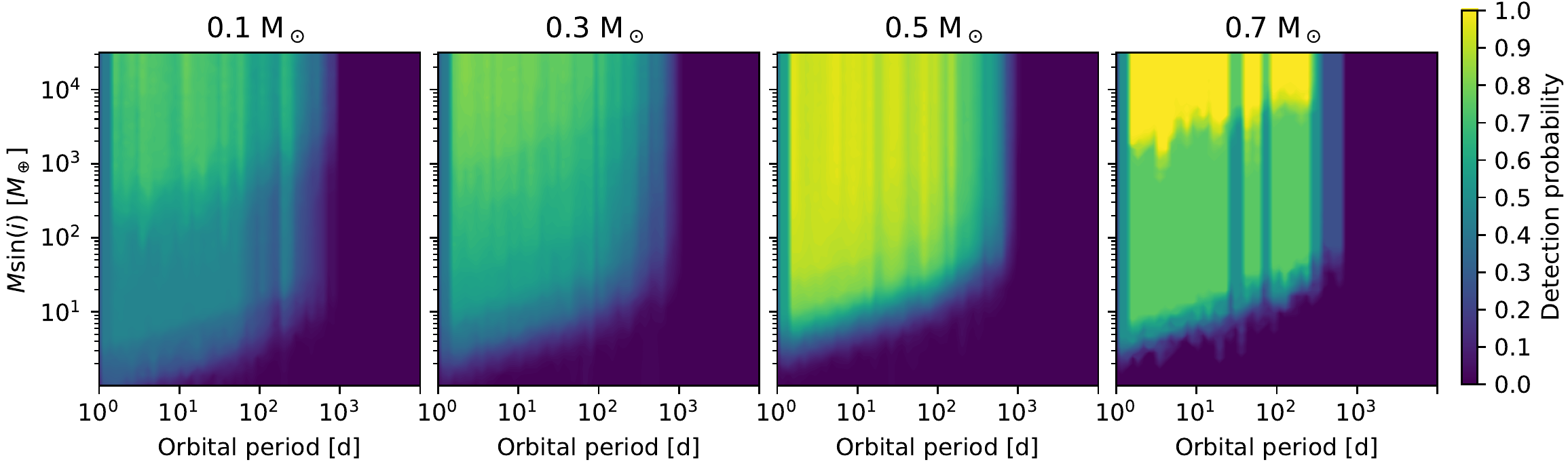}
    \caption{Combined detection probabilities for the HARPS\&CARM$_{70}$ survey binned to the stellar masses  used in the simulations.
    The color code represents the detection probability of a planet with the corresponding period, minimum mass, and host star mass.
    The zero-detection threshold resides at slightly lower minimum masses for lower-mass stars, and the detection map for \SI{0.7}{\Msun} stars is patchy due to the small sample there.}
    \label{fig:sensitivities}
\end{figure*}

To make sure that the analysis is as self-consistent as possible, we computed survey sensitivities for the HARPS survey~\citep{Bonfils2013} in the same way using RV measurements and activity indicators produced with~SERVAL \citep{Zechmeister2018} by \citet{Trifonov2020}.
To produce global sensitivity maps for a combined HARPS\&CARM$_{70}$ survey, we averaged the detection probability of each grid point for all targets of the combined sample.
The resulting maps for the four stellar mass bins are shown in Fig.~\ref{fig:sensitivities}.

Each \textit{NGM} planet received a detection probability depending on the period-minimum mass bin in which it falls on this map.
We then assigned each synthetic planet a uniformly drawn random number from the interval $[0.0, 1.0)$ and considered the planet detected if this number is lower than the planet's detection probability.
Out of the \NplanetsNGMzero\ synthetic planets, only \Ndetectable\ are detectable according to this procedure.

\section{Results}\label{sec:results}

\subsection{Observed and theoretical planet detections}
Occurrence rates are often computed from incomplete catalogs by characterizing a detection bias and correcting for it~\citep[e.g.,][]{Cumming2008,Bonfils2013,Burke2015,Obermeier2016,Lienhard2020,Wittenmyer2020a,Gibbs2020}.
This has the disadvantage of having to extrapolate to domains of low sensitivity, leading to large uncertainties~\citep{Foreman-mackey2014}.
Instead, here we apply the calculated bias of the combined HARPS\&CARM$_{70}$ survey to our synthetic population and directly compare the observed sample with detectable synthetic planets.
We focus on comparing planetary detection rates $n$, which refer to the number of simulated or observed planet detections per finite interval in a parameter (e.g., the orbital period).
Owing to the RV-detected exoplanet sample, our parameter space of choice is the one spanned by the minimum mass $\msini$ and the orbital period $P$.
The corresponding detection rate density in this parameter space can be expressed as
\begin{equation}
	\Gamma_{M,P} = \frac{\partial^2n}{\partial \log \msini \, \partial \log P}.
\end{equation}

Across all masses and periods, we find a total number of observed planets per star of $\ntotobs \pm \ntotobserr$ in the HARPS\&CARM$_{70}$ sample and a total number of observable synthetic planets per star of $\ntotsyn \pm \ntotsynerr$, where the uncertainties are Poisson errors from the counting statistic.
With \Nplanets\ planets in \Nplanetssystems\ systems, the observed mean multiplicity of \multPlanets\ is in agreement with the synthetic one of \multDetectable\ (\Ndetectable\ synthetic planets in \Ndetectablesystems\ systems).
The total observed and theoretical detection rates thus appear consistent with each other.
\rev{We note that the detection rate of giant planets in the CARMENES sample could be overestimated since the survey is not yet completed~(see Sect.~\ref{sec:caveats}, \citet{Sabotta2021}).}

\begin{figure}
	\centering
	\includegraphics[width=\hsize]{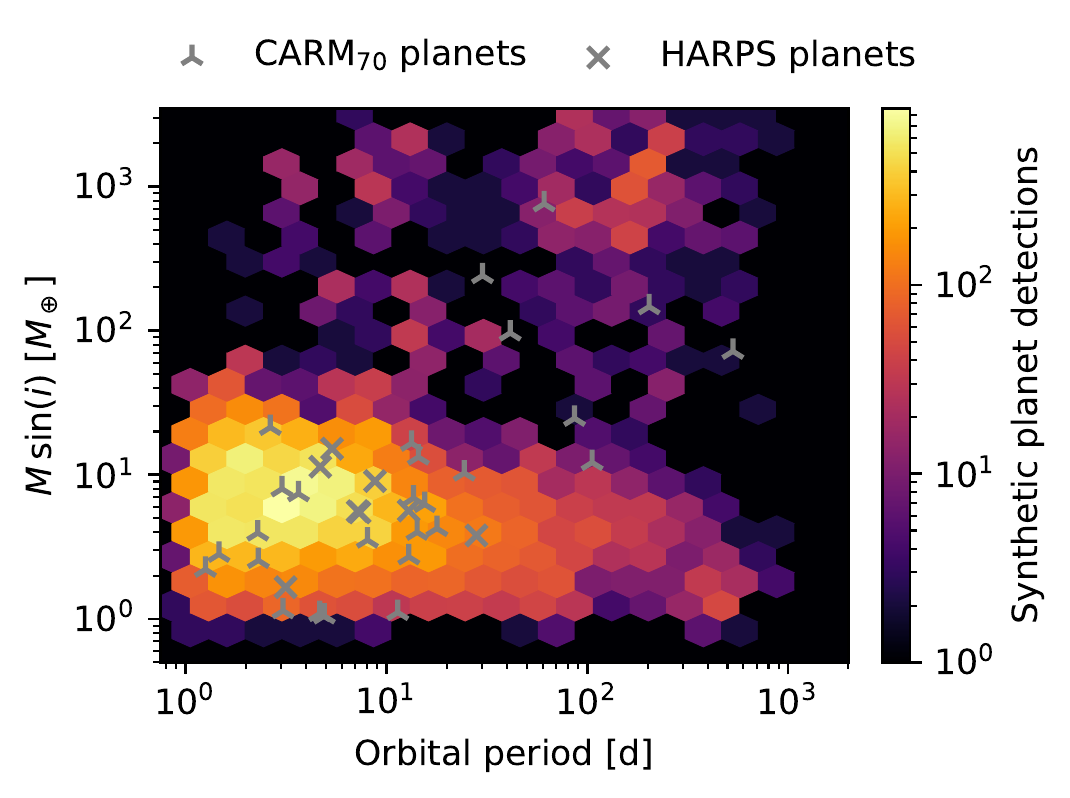}
	\caption[]{Observed and biased synthetic planets in minimum mass-period space. Markers show confirmed planets from the HARPS and CARM$_{70}$ surveys, respectively. The color code shows the frequency of simulated planets in this parameter space after applying the detection bias of the combined HARPS\&CARM$_{70}$ survey.
    Low-mass planets are well reproduced by the model, but intermediate-mass planets appear more common than theoretically predicted. 
	}
	\label{fig:msini-period_hex}
\end{figure}
Figure~\ref{fig:msini-period_hex} shows the distribution of biased simulated and observed planets in minimum mass-period space.
The bulk of the unbiased theoretical occurrence density lies at roughly Earth-mass planets at orbital periods of a few hundred days~\citep{Burn2021}.
After applying the observational \rev{bias,} the peak density is shifted to the super-Earth mass regime with periods of a few days.
\rev{Super-Neptunes and giant planets ($\msini \gtrsim \SI{20}{\Mearth}$) are expected to be much rarer.
The model anticipates detections to be most likely at $\msini \gtrsim \SI{200}{\Mearth}$ with orbits around \SI{\sim 100}{\day}.
It produces the fewest planets in the \SIrange{20}{200}{\Mearth} mass range, especially in orbits beyond \SI{\sim 20}{\day}.}
\rev{However, such a dearth of Neptune to Saturn-mass planets at intermediate orbits} is not obvious in the sample of detected CARM$_{70}$ and HARPS planets.
Although these are subject to selection effects (see Sect.~\ref{sec:caveats}), it is noticeable that the mass distribution here is rather continuous and does not reveal multiple modes.
Concerning the agreement of simulated and observed planets, low-mass planets up to \rev{around \SI{10}{\Mearth}} are well reproduced by the model, whereas intermediate-mass planets occur in the observed sample but are hardly present in the synthetic population.

\subsection{Planet detections as a function of host star mass}
\begin{figure*}
    \centering
    \includegraphics[width=\textwidth]{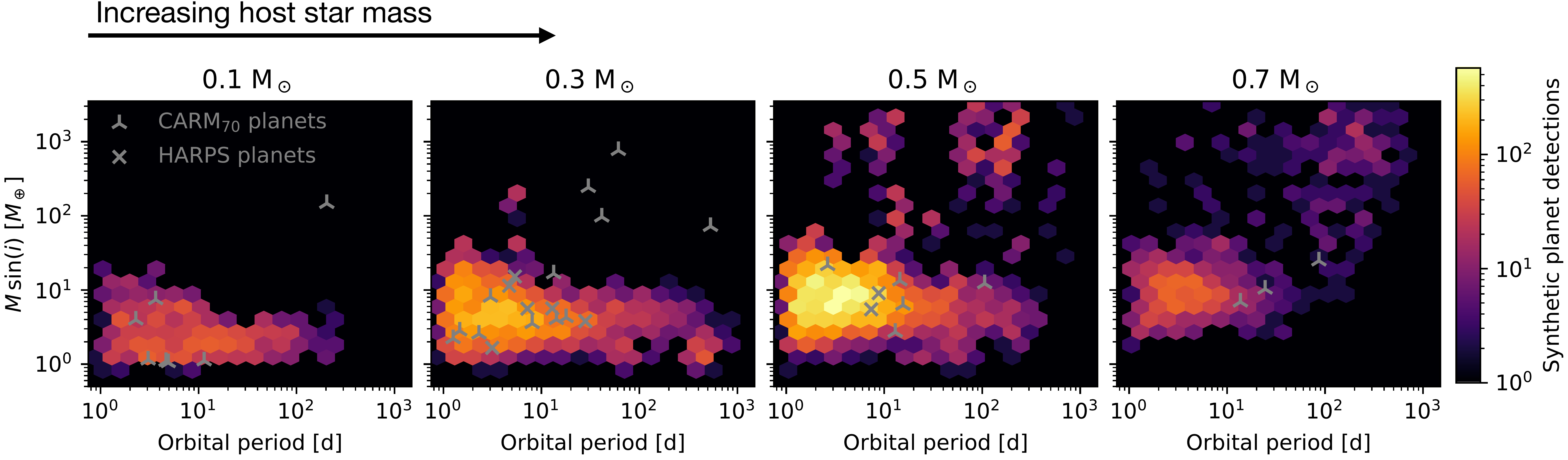}
    \caption[]{As Fig.~\ref{fig:msini-period_hex}, but partitioned according to host star mass.
    In the biased synthetic population, the domination by low-mass planets on short to intermediate orbits is similar for all stellar masses, but giant planets occur only around stars with masses \SI{0.5}{\Msun} and higher.
    In contrast, the subgiant and giant planets in the observed sample occur most frequently around less massive stars.
    GJ~3512b, the giant planet in the \SI{0.1}{\Msun} panel, is particularly at odds with theoretical predictions. No simulated planets with $M \sin (i) \gtrsim \SI{10}{\Mearth}$ occur in this stellar mass bin.
    }
    \label{fig:quad_msini-period_hex}
\end{figure*}

To identify trends with planet host mass, we explore planet detections in the four stellar mass bins defined above~(Fig.~\ref{fig:quad_msini-period_hex}).
The synthetic population shows a clear trend of increasing giant planet detection rate with host star mass.
In general, giant planets are rare and occur only in the stellar mass bins $\SI{0.5}{\Msun}$ and $\SI{0.7}{\Msun}$.
However, the giant planets in the HARPS\&CARM$_{70}$ sample are in stark contrast to this trend.
The observed giant planet closest to a synthetic counterpart is \object{GJ 876b}, which orbits its host with a period of \SI{61}{\day} and has a projected mass of $\msini \approx \SI{761}{\mEarth}$~\citep{Marcy2001,Rivera2005,Trifonov2018}.
With a stellar mass of \SI{0.37}{\Msun}, the system just barely ended up in the stellar mass bin with zero detection rate density in the giant domain and is in fact relatively close to the few synthetic giant planets in the \SI{0.5}{\Msun} population.

There are four discovered planets on intermediate and large orbits ($P = \SIrange{10}{1000}{\day}$) with projected masses $\msini = \SIrange{20}{200}{\mEarth}$, where \textit{NGM} shows a deep valley in the detection rates: \object{GJ 1148b},c~\citep{Haghighipour2010,Trifonov2018}, \object{HD 147379b}~\citep{Reiners2018a}, and \object{GJ 3512b}~\citep{Morales2019}. 
While none of them would have been expected based on our simulations, the \SI{147}{\mEarth} giant \object{GJ 3512b} is particularly difficult to reconcile with theoretical predictions: it orbits a late (M5.5) M~dwarf with a very low mass of \SI{0.123\pm0.009}{\Msun}~\citep{Morales2019}.
Our theoretical model produces no giant planets in this stellar mass regime, and its existence remains a challenge for planet formation theories based on core accretion~(\citealp[][but also see]{Liu2020,Burn2021,Schib2021} \citealp{Kurtovic2021}).

On the other hand, low-mass planets ($\msini \lesssim \SI{30}{\Mearth}$) are well represented by our synthetic systems.
Populations from all stellar mass bins contribute to the synthetic planet detections in this domain, where in the unbiased population the maximum occurence rate density is invariably at orbits \SIrange{\sim100}{1000}{\day} and terrestrial planet masses of a few \SI{}{\mEarth}.
In the biased population presented here, this peak is shifted toward periods on the order of a few days.
The observed low-mass planets in CARM$_{70}$ and HARPS lie all in the domain where significant synthetic detection rate density exists.

\begin{figure}
	\centering
	\includegraphics[width=\hsize]{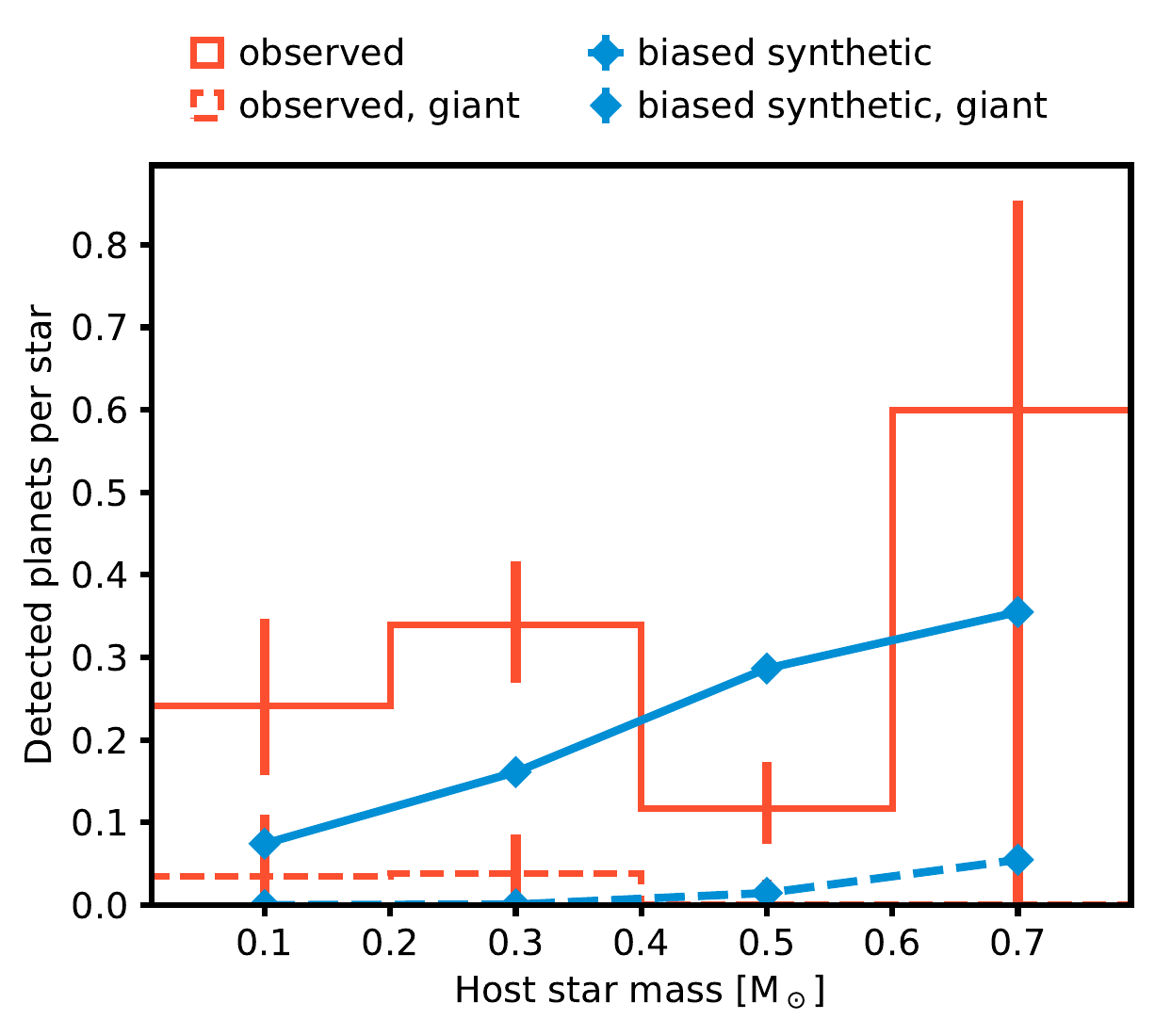}
	\caption[]{Simulated and observed planet detections as a function of host star mass. In each stellar mass bin, vertical lines denote \rev{\SI{68}{\percent} confidence intervals based on the binomial distribution (not visible for the synthetic population),} and the counts are normalized to the number of stars in that bin. The synthetic population features discrete $M_\star$ values and shows a linear increase of detections with increasing stellar mass.
    The observed sample shows \rev{no clear trend.}
	}
	\label{fig:hist_detections_Mstar}
\end{figure}
If we normalize the planet detections by the number of stars of a given host star mass range (see Fig.~\ref{fig:hist_detections_Mstar}), the simulated detections increase linearly with stellar mass.
\rev{This can be directly attributed to the scaling of disk masses (see Sect.~\ref{sec:BernModel_mdwarf}).}
In contrast, \rev{the observed detections} vary across the considered range without a clear trend.
\rev{While the total numbers of detections in both samples agree, significant deviations exist when they are partitioned by stellar mass.
This is particularly true for giant planets.}

\subsection{Planetary mass function}
By marginalizing over the orbital period axis, a minimum mass distribution of the samples can be obtained.
\begin{figure}
    \includegraphics[width=\hsize]{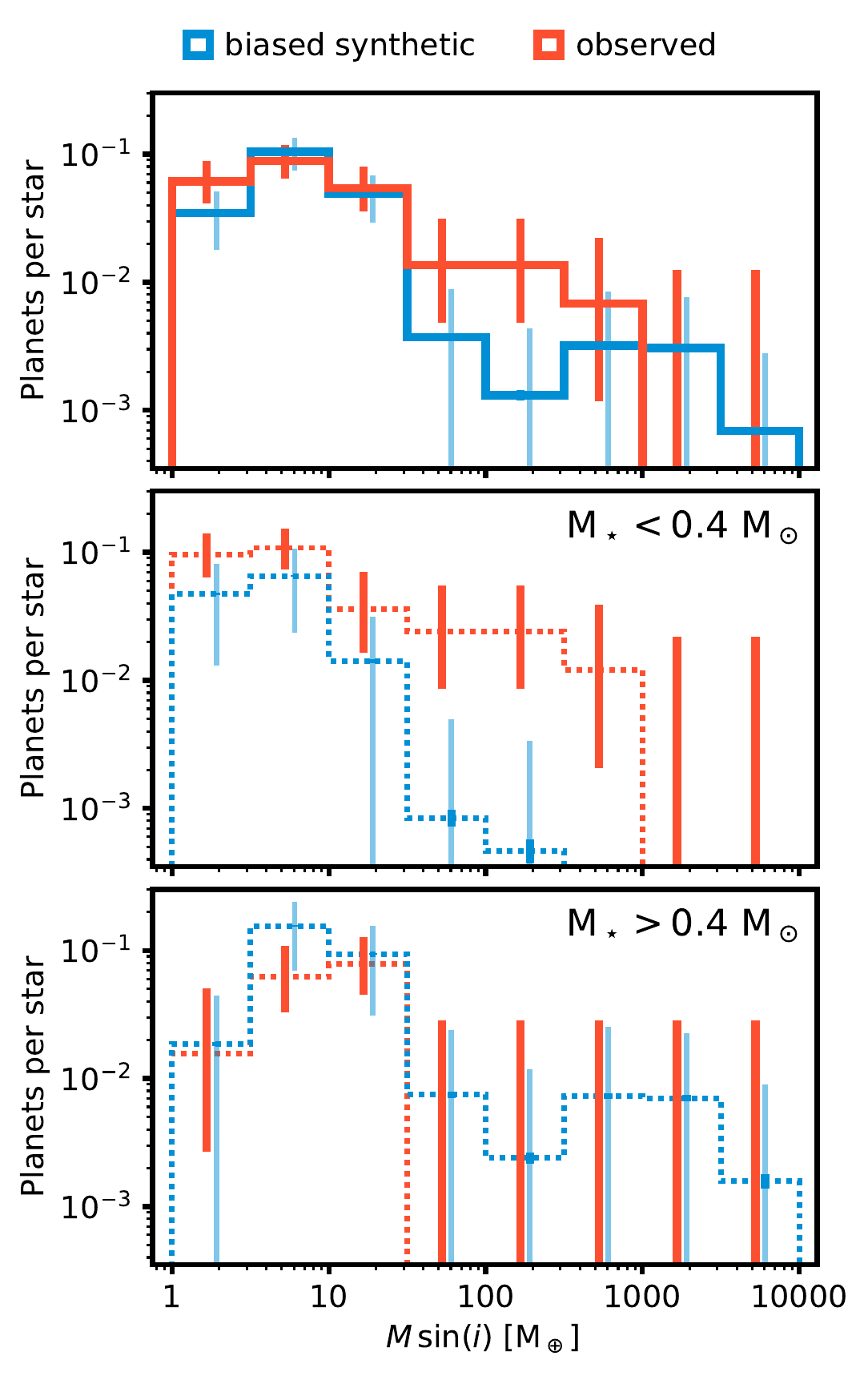}
    \caption[]{Minimum mass distribution of synthetic and observed planets.
    Thick vertical lines show \rev{\SI{68}{\percent} confidence intervals based on the binomial distribution.}
    Thin blue lines show standard deviations of 2000 bootstrapped synthetic samples, where each sample had the size of the observed sample.

    Top: Total normalized counts.
    The distribution of the biased synthetic population appears bimodal in $\msini$.
    The observed sample shows a peak in the terrestrial planet regime of a few Earth masses and a continuous downward slope without a valley.
    At low masses, the theoretical and observed distributions agree.
    The formation model underproduces planets $\gtrsim \SI{30}{\Mearth}$ and features a ``sub-Saturn valley\rev{.}''

    Center and bottom: Normalized counts for late ($<\Mthresh \,\SI{}{\mSun}$) and early ($>\Mthresh \,\SI{}{\mSun}$) M~dwarfs separately.
    At minimum masses beyond $\sim \SI{30}{\Mearth}$, theory and observations disagree:
    while observed subgiant and giant planets occur mostly around stars with masses $\sim\SI{0.3}{\mSun}$, the formation model produces such planets only around more massive stars.
    }
    \label{fig:Msini_hist_Mstar}
\end{figure}
Figure~\ref{fig:Msini_hist_Mstar} shows normalized histograms in $\msini$ for the observed and simulated planet samples.
In the domain of terrestrial planets and super-Earths up to $\sim \SI{30}{\Mearth}$, the observed and theoretical distributions match well.
Beyond that, the model predicts \rev{fewer} planets than observed and shows a few very massive planets ($> \SI{1000}{\Mearth}$) that are not observed.
To quantify the difference between the observed and theoretical planetary mass functions, we performed a two-sample Anderson-Darling test~\citep{Anderson1952}, which is based on a nonparametric and distribution-free test statistic $A^2$.
Using the implementation in \textit{Scipy}\footnote{\url{https://docs.scipy.org/doc/scipy/reference/generated/scipy.stats.anderson_ksamp.html}} and critical values from \citet{Scholz2007}, we find that the null hypothesis that the two samples stem from the same distribution can only be rejected at an estimated \SI{25}{\percent} significance level ($A^2 = \adstatisticMsini$).
We cannot \rev{exclude} an identical underlying base distribution.
If we divide the samples into early ($>\Mthresh \,\SI{}{\mSun}$) and late ($<\Mthresh \,\SI{}{\mSun}$) host stars\footnote{$\Mthresh \,\SI{}{\mSun}$ is centered between two of our simulated, discrete stellar masses and close to the threshold mass where stars become fully convective~\citep{Cifuentes2020}.}, the agreement remains only for the early subsample.
Due to the large difference in giant planet detection rates, the observed and simulated samples for later spectral subtypes are different at the \SI{~0.1}{\percent} level.

Figure~\ref{fig:Msini_hist_Mstar} further shows an apparent bimodality of the synthetic $\msini$ distribution for early host stars.
We tested the significance of such a bimodality by applying Hartigan's dip test~\citep{Hartigan1985}, which tests the null hypothesis of a unimodal distribution, on the different synthetic subsamples.
For the overall synthetic sample, the resulting p-value of $\dipPvalNGM$ suggests a multimodal distribution.
Distinguishing between early and late stars, the dip test suggests a multimodal distribution for the early sample ($p_\mathrm{early} = \dipPvalNGMEarly$) but not for the late sample ($p_\mathrm{late} = \dipPvalNGMLate$).
This is because at stellar masses below \Mthresh $\,\SI{}{\mSun}$ no giant planets occur and the distribution features a single slope.
The minimum mass distribution of the observed sample is overall more flat and shows an almost continuous negative slope.
Consequently, Hartigan's dip test suggests a unimodal distribution ($p_\mathrm{obs} = \dipPvalobs$).

\subsection{Orbital period distributions}\label{sec:results_orbital_periods}
\revii{One of the most directly accessible} planetary properties is the orbital period.
\begin{figure}
    \includegraphics[width=.99\hsize]{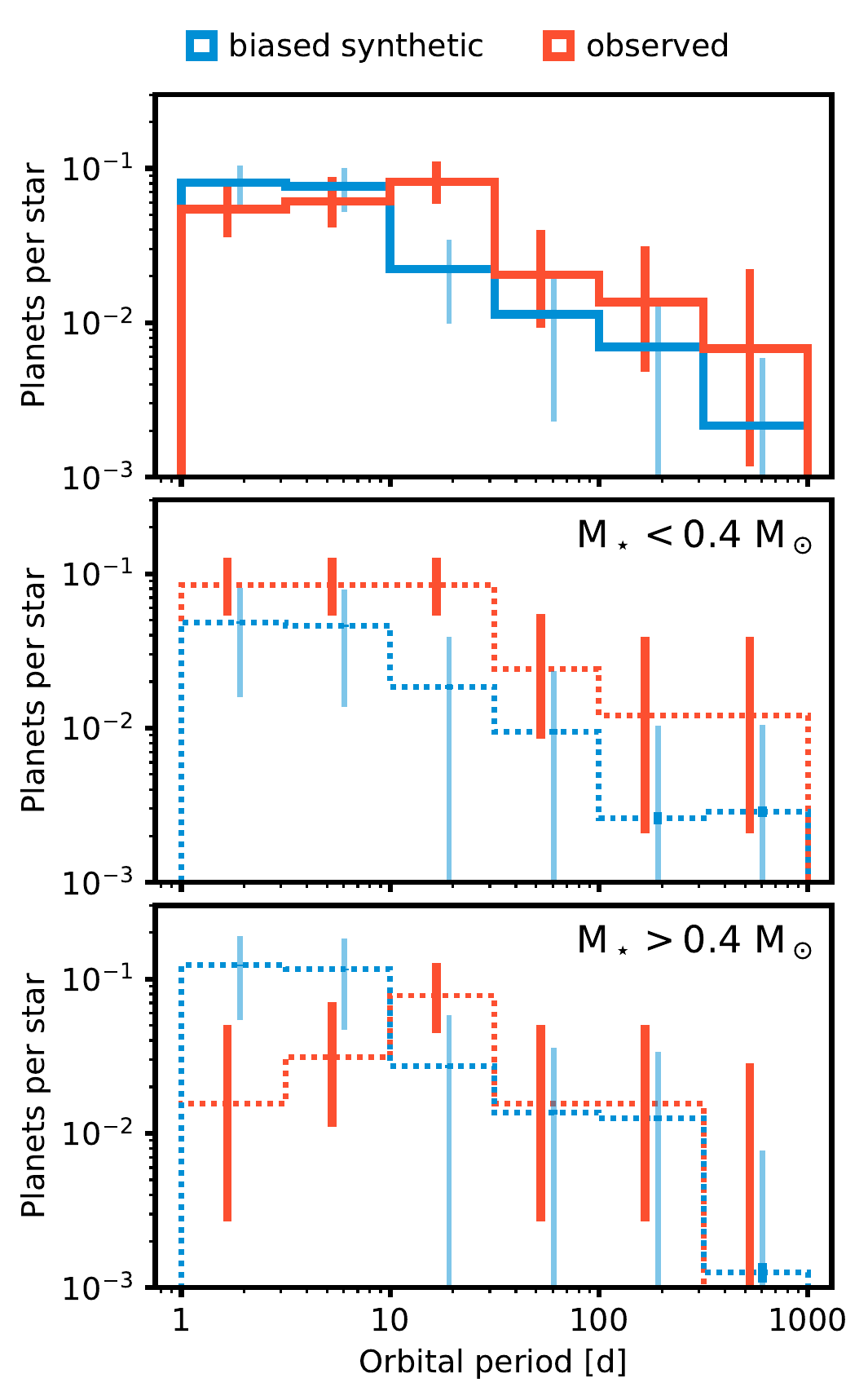}
    \caption[]{
       Period distribution of observed and synthetic planets. As in Fig.~\ref{fig:Msini_hist_Mstar}, thick error bars denote \rev{\SI{68}{\percent} confidence intervals based on the binomial distribution,} and thin lines show the standard deviations of bootstrapped simulated samples.

Top: Total normalized counts.
        The distributions are comparable over most of the range, except for a peak at \SIrange{\sim10}{30}{\day} that only occurs in the observed sample.

Center and bottom: Normalized counts for early ($>\Mthresh \,\SI{}{\mSun}$) and late ($<\Mthresh \,\SI{}{\mSun}$) M~dwarfs separately, both showing a high detection rate at a few tens of days in the observed sample.
Significantly fewer short-period planets are found around higher-mass stars, which is in disagreement with our model.
    }
    \label{fig:period_hist_Mstar}
\end{figure}
The period distributions of the synthetic and observed planets agree well, except for a peak at a few tens of days that \rev{occurs only }in the observed sample (Fig.~\ref{fig:period_hist_Mstar}).
\revii{Orbital periods of the biased synthetic population resemble a power~law with a single slope, and a} common origin of the distributions can be excluded on the \SI{0.1}{\percent} level ($A^2 = \adstatisticPeriods$).
\revii{The apparent log-linear slope is partly shaped by the period-sensitive completeness function, which masks a slight drop in occurrence for planets with periods shorter than \SI{\sim 10}{\day}.}

\revii{The observed sample shows} a dearth of \revii{such short-period planets} around stars of higher masses.
This feature is not reproduced in our model with its current prescription for planet migration and the innermost parts of the protoplanetary disk~(see discussion in Sect.~\ref{sec:discussion_periods}).
\todo[inline]{multimodal period dist in single vs. multis: \citep{Schlecker2021b}} 

\section{Discussion}\label{sec:discussion}
The overall number of planet detections per star in the observed HARPS\&CARM$_{70}$ sample and in the biased synthetic \textit{NGM} population are in good agreement.
In $\msini$-$P$-$M_\star$ space, some discrepancies occur that we discuss here.

\subsection{A missing sub-Saturn valley in the planetary mass function}
The simulated and observed minimum mass distributions agree for small planets up to \SI{\sim 30}{\Mearth}.
Beyond that, their shapes deviate:
while the minimum mass distribution recovered from the discovered planets follows a smooth power law, the simulated planets show a significant \rev{bimodality.
The valley between about \SI{30}{\mEarth} and \SI{200}{\mEarth} is also present in the unbiased synthetic population~\citep{Burn2021}.}
Testing the existence or nonexistence of this demographic feature in M~dwarf planetary systems may provide clues about the gas accretion process \rev{in the core accretion paradigm for different stellar types.}

In the classical picture, the valley separates all planets that attained solid cores massive enough to enter runaway gas accretion and became giant planets from those that did not~\citep{Mizuno1978,Pollack1996}.
Due to the short duration of the runaway phase, only few planets retain intermediate masses~\citep[][Emsenhuber et al., in prep.]{Mordasini2009b,Emsenhuber2021b}.
Three-dimensional hydrodynamical simulations of planetary gas accretion have challenged this prediction by proposing significantly lower accretion rates~\citep{Szulagyi2014,Moldenhauer2021}.

Preliminary observational evidence for the predicted valley has been provided by \citet{Mayor2011}, who computed bias-corrected occurrence rates and the mass distribution for the HARPS RV survey, albeit for mostly solar-type stars.
They reported a decrease of their bias-corrected mass distribution ``between a few Earth masses and $\sim \SI{40}{\Mearth}$.''
In contrast, the valley appears absent in the MOA-II microlensing survey~\citep{Suzuki2016,Suzuki2018}.
Furthermore, the significance of the \citet{Mayor2011}-dip has recently been questioned~\citep{Bennett2021}.

Future expansion of the observational sample will shed light on the existence, strength, and physical origin of a ``sub-Saturn valley.''
Planets on intermediate orbits of \SIrange{\sim 10}{100}{\day} period~\citep[e.g.,][]{Espinoza2016g,Kipping2019,Brahm2020,Schlecker2020,Hobson2021} are suitable study objects, as they are less affected by direct interaction with the host star than their hotter siblings~\citep{Thorngren2016}.
The absence of a significant valley in our sample already indicates that it is not as pronounced as our canonical gas accretion formalism predicts.

\subsection{\rev{Excess of giant planets around late M~dwarfs}}
The stellar mass \rev{dependence of giant planet detections} with $\msini > \SI{100}{\Mearth}$ in the synthetic \textit{NGM} and observed samples are at odds with each other.
While the formation model generally produces giant planets only around earlier stars, our RV-detected giant planets orbit only host stars with masses lower than \SI{0.5}{\mSun}.
This is despite a significantly higher survey sensitivity around earlier stars~\citep{Sabotta2021}.

Further constraints on the occurrence of giant planets \rev{around stars over a wide range of masses} are provided by the California Legacy Survey \rev{\citep[CLS,][]{Rosenthal2021,Fulton2021}}, which combined and extended previous surveys using Keck and HIRES data \citep{Cumming2008,Howard2010a,Hirsch2021}.
\begin{figure}
    \centering
    \rev{\includegraphics[width=\linewidth]{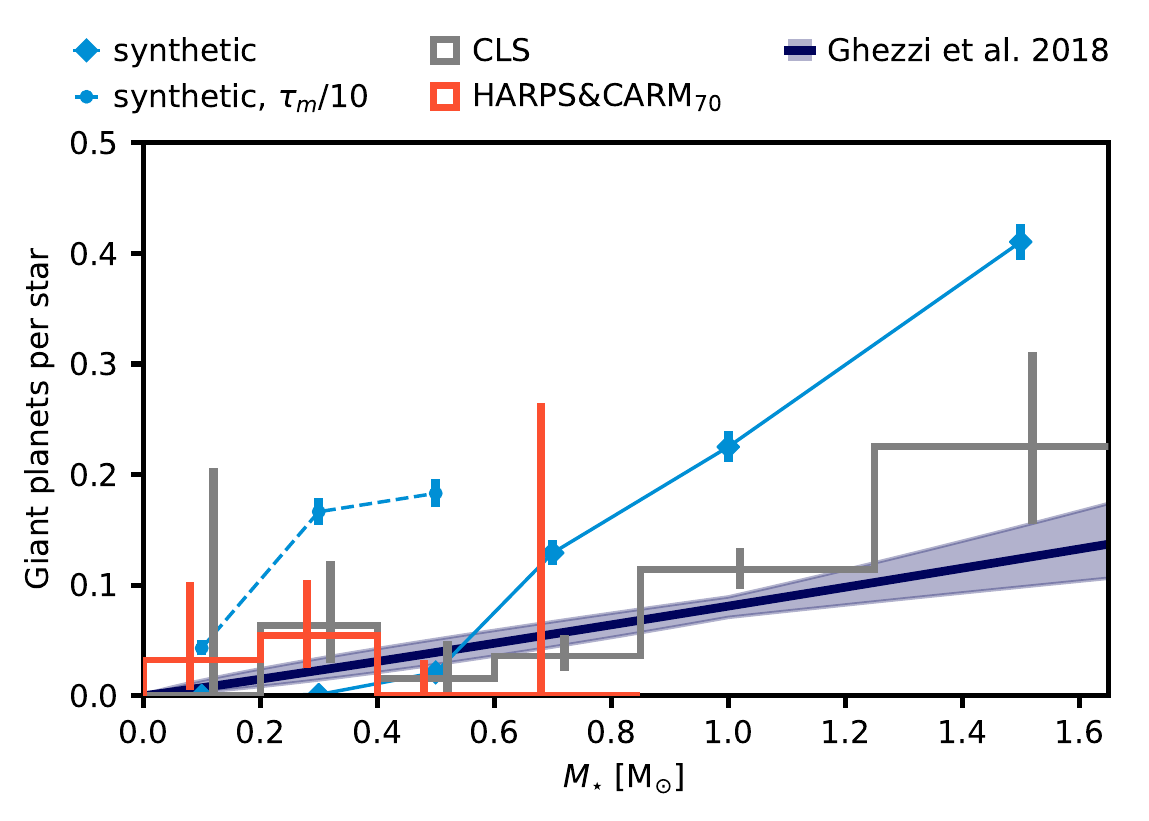}}
    \caption{Giant planet detections as a function of stellar mass.
    From the HARPS and CARMENES planets presented here (red), the California Legacy Survey \citep[][gray]{Rosenthal2021}, and our synthetic \textit{NGM} planets (blue), we include detections with RV semi-amplitude $K>\SI{10}{\meter\per\second}$, $\msini>\SI{100}{M_\oplus}$, and $P<\SI{1000}{d}$.
    We removed planets around close spectroscopic binaries ($P<\SI{20}{yr}$) reported in \citet{Rosenthal2021}.
    \rev{In addition, we show the fit of \citet{Ghezzi2018} for metallicities consistent with the assumptions made to draw the synthetic planets.
    Shaded areas and vertical lines denote \SI{68}{\percent} confidence intervals, respectively.}
    We note that the HARPS\&CARM$_{70}$ sample is limited to M~dwarfs $\lesssim\SI{0.7}{M_{\odot}}$.
    \rev{The observed giant planets around very low-mass stars indicate a possible break of the monotonic stellar mass trend at around $\sim$\SI{0.5}{M_{\odot}},} which theoretical predictions fail to reproduce.
   If \mbox{type-I} migration is inhibited by a factor $\times 10$ in the simulations (dashed blue line), giant planets occur also in systems with lower stellar masses.}
    \label{fig:CPS_giants}
\end{figure}
We plot its giant planet detections as a function of the stellar host mass in Fig.~\ref{fig:CPS_giants} together with \rev{the combined HARPS\&CARM$_{70}$ sample and our synthetic systems, where we included analogous populations for \SI{1.0}{\Msun} and \SI{1.5}{\Msun} stellar hosts~\citep{Emsenhuber2021b}.
    We further add the linear giant planet occurrence trend found in~\citet{Ghezzi2018}.
For the latter, we use the stellar mass-metallicity fit of their most recent planetary sample at $\mathrm{[Fe/H]}=-0.02$, which is the mean metallicity in the synthetic sample of \citet{Burn2021}. 
The fit deviates from the CLS sample for larger stellar masses due to enhanced metallicities in the latter.
Thus, the derived, almost linear dependency ($\propto M_\star^{1.05}$) by \citet{Ghezzi2018} is better suited for comparison to the synthetic data than the full CLS sample.
Similar to previous findings~\citep{Kennedy2008}, our synthetic giant planet occurrences \revii{show a trend} in stellar mass that is significantly steeper than the observed one.
This suggests that the gas accretion process needs to be revised in the model (Emsenhuber et al., in prep).}

\rev{Instead of changing the formation process itself, the proposed stellar mass scaling of the initial disk mass, which controls the number and masses of planetary cores~\citep[e.g.,][]{Kennedy2008}, may be questioned.
It is based on millimeter dust continuum emission, which may be affected by unknown opacities, conversion of solid material into larger particles, or optical depth effects~\citep[e.g.,][]{Molyarova2017,Pascucci2016}.
    A shallower stellar mass-scaling could resolve the discrepancy, however, realistic stellar accretion rates should be maintained.
In our nominal scaling, these are consistent with observed data by \citet{Alcala2017} \citep[see][]{Burn2021}, although slightly shallower.
Indeed, other authors matched observed accretion rates with a steeper than linear slope: \citet{Alibert2011} found a good match to observational data \citep{Muzerolle2003,Natta2004} for a slope of $1.2$.
Similarly, the scaling of \citet{Manara2012} used in \citet{Liu2020} results in a steeper relationship.
Thus, a shallower than linear slope for the disk mass as a function of stellar mass is disfavored by accretion rate observations and not a viable option to enhance the giant planet formation rate around low-mass stars.}

Giant planets occur more frequently around stars of higher metallicity~\citep{Gonzalez1997,Santos2001,Santos2004,Fischer2005,Buchhave2018}.
Since for a fixed disk mass, metallicity controls the solid budget of the disk, larger metallicities promote the formation and growth of planetesimals and ultimately of planetary cores able to reach the runaway regime.
It is thus reasonable to suspect that \rev{the giant planets in our sample} have formed in particularly metal-rich environments that are not represented by the metallicity distribution of the synthetic systems.
The reported values for the giant planet hosts mentioned above are $0.05 \pm 0.2$\,dex or $-0.11 \pm 0.10$ for Gl 876 b \citep{Correia2010,Marfil2021}, $0.36\pm0.2$ for Gl 317 \citep{Anglada-Escude2012}, $0.13\pm 0.16$ or $-0.17 \pm 0.10$ for GJ 1148 \citep{Passegger2018,Marfil2021} and $-0.07\pm0.16$ or $-0.12\pm0.16$ for GJ 3512 \citep{Morales2019,Marfil2021}, leaving only Gl~317 with an enhanced metallicity.
Although the uncertainties are large, \rev{the} other targets are consistent with solar metallicity.
\rev{Recent discoveries in the literature like the sub-Neptune around TOI-2406 ($M_\star = \SI{0.16}{M_{\odot}}$, \mbox{[Fe/H]} = \SI{-0.38\pm0.07}{},~\citet{Wells2021}) and the giant planet around a late \mbox{(M5.0~V)} M~dwarf with subsolar metallicity~\citep{Quirrenbach2022} complement the pattern.}
A high metal content alone thus cannot explain the existence of giant planets around very low-mass stars.

\rev{Already \citet{Johnson2010} reported an elevated giant planet occurrence at stellar masses below \SI{0.5}{\Msun}, and both CARMENES and CLS find a slight} excess of giant planets around stars with masses $\sim \SI{0.4}{M_{\odot}}$ \rev{compared to the linear trend in~\citep{Ghezzi2018}:}
\object{Gl 876b} and c (J22532-142, \citealp{Marcy1998,Delfosse1998,Marcy2001,Rivera2005,Millholland2018,Trifonov2018}) are included in both the CLS and the CARMENES sample.
Furthermore, \object{Gl 317} \citep{Johnson2007,Anglada-Escude2012} and \object{HIP 57050} (GJ~1148, Ross~1003, see also \citealp{Trifonov2020a}) both host a planet with mass \SI{>100}{\Mearth} and are included in CLS but not in the CARM$_{70}$ sample (the latter is reported with a mass $\msini$\SI{<100}{\Mearth} in CARMENES).

\rev{Stellar masses below \SI{\sim 0.3}{\Msun} have not been thoroughly explored by RV surveys, and the currently small sample does not allow us to draw a definite conclusion about a deviation from the linear trend at very low stellar masses.
However, \citet{Quirrenbach2022} recently reported the discovery of another giant planet around a late \mbox{(M5.0~V)} M~dwarf.
Although not part of our sample, the roughly Saturn-mass planet provides additional evidence for the existence of giant planets around stars of all masses.}

\rev{It also allows a simple estimate:
For 66 stars of spectral type \mbox{M4.0~V} or later, which roughly corresponds to our lowest-mass bin, CARMENES has already collected 30 or more RV measurements.
This rather arbitrary threshold serves as a reasonable reference value above which CARMENES would detect gas giants in a wide range of periods.
Two of these stars have been shown to host giant planets~\citep{Morales2019,Quirrenbach2022}.
We can thus estimate a lower limit on the giant planet occurrence rate of $2/66 \approx \SI{3}{\percent}$ in this stellar mass bin.
This assumes a generic completeness of \SI{100}{\percent} -- the actual occurrence rate is likely higher when realistic detection sensitivities are taken into account.
This crude estimate hints at a possible break of the linear occurrence rate trend reported before.
However, no conclusions in this regard should be drawn ahead of a thorough sensitivity and occurrence rate assessment upon completion of the survey.}

\rev{Discoveries} of gas giants through gravitational microlensing support the existence of massive planets around low-mass stars~\citep[e.g.,][]{Udalski2005,Dong2009,Bennett2020,Bhattacharya2021,Han2021}.
These \rev{discoveries} reinforce the puzzling disagreement of the new CARMENES results with theoretical predictions of giant planet occurrence as a function of stellar mass.
\rev{This raises the question of whether} planets with extreme planet-to-star mass ratios may form within the core accretion framework, or, as hypothesized in~\citet{Morales2019}, via a disk instability scenario~\citep{Cameron1978}.

\subsection{\rev{Potential of core accretion to produce giant planets around low-mass stars}}
\rev{There is a substantial body of literature that provides predictions about the existence of gas giant planets around low-mass stars.}
From the point of view of disk and exoplanet observations,~\citet{Manara2018} found that the combined solid masses of planetary systems frequently exceed those of the most massive dust disks\rev{ ($\SI{30}{\mEarth} \approx \SI{e-4}{\mSun}$ for M~dwarfs), leading them to speculate about} a dedicated formation pathway for massive giant planets around very low-mass stars.
\rev{It should be noted that the total mass and distribution of planetesimals may strongly deviate from what is implied by the gas and dust distribution in disks~\citep{Lenz2019,Voelkel2020}.
At the times of observation, a large fraction of the dust might already be hidden in planetesimals~\citep{Gerbig2019}.}

\citet{Miguel2020} performed a population synthesis of planetary systems around (very) low-mass stars ($M_\star = \SIrange{0.05}{0.25}{\Msun}$) based on a semi-analytical model assuming classical planetesimal accretion.
Their model was originally designed to study circumplanetary disks and features a low-viscosity gas disk model as well as \mbox{type-I} and \mbox{type-II} migration~\citep{Miguel2016}.
They find efficient planet formation only in sufficiently massive disks ($\gtrsim \SI{e-2}{\mSun}$).
Even under these conditions, their model fails to form any planets more massive than~\SI{5}{\mEarth}.

\rev{Using a pebble accretion based model, \citet{Liu2019} model gas accretion onto planetary cores that have reached pebble isolation mass~\citep{Lambrechts2014}. Their gas accretion method is optimistic (reduced opacity by a factor of $\sim 20-100$, no leftover thermal energy due to prior solid accretion) and therefore useful as an upper limit for giant planet formation in the pebble accretion scenario \citep[see also][]{Brugger2020}.
In this case and for varied disk properties, they found a minimum stellar mass for giant planet formation of \SI{\sim 0.3}{M_{\odot}}.}

\rev{In good agreement, \citet{Mulders2021} also use a pebble drift and accretion model to simulate planet growth around stars with different masses. They find that no giant planets form at stellar host masses $\lesssim \SI{0.3}{\Msun}$.
}

\rev{\citet{Adams2021} explore the planetary mass function of planets formed around host stars of different masses with a semi-empirical approach.
They find a steeper planetary mass function for low-mass stars, suggesting a low probability of producing Jovian planets around stars with $M_\star \lesssim \SI{0.5}{\Msun}$.
}

\citet{Zawadzki2021} used \mbox{\textit{N}-body} calculations to simulate the formation of planets around \SI{0.2}{\mSun} stars.
Under the assumptions of early planetesimal formation~\citep{Lenz2019} and including \mbox{type-I} migration, they find efficient growth of planetary cores through early collisions of planetary seeds.
Their setup \rev{assumes a rather high solid disk mass} of~$\sim \SI{2e-2}{\Msun}$ at the start of their simulations.
While they do not model gas accretion onto planets, many of these cores grow to super-Earths, some of which in the mass range where runaway accretion could be triggered.

Finally, \citet{Burn2021} \rev{discuss modifications} to the model used here that would enable the formation of giant planets around very low-mass stars.
The dashed line in Fig. \ref{fig:CPS_giants} shows the fraction of giant planets in simulations with \mbox{type-I} migration velocities inhibited by a factor~$\times 10$.
For computational cost considerations, we initialized these otherwise identical simulations with only 20 planetary cores per disk, which is adequate for modeling gas giant systems~\citep{Emsenhuber2021b}.
Suppressing planet migration clearly has a strong effect on giant planet formation:
In the modified simulations, planetary cores are able to reach masses beyond \SI{\sim 10}{\mEarth} even around late M stars without rapidly migrating into the star, enabling giant planet formation.

Even though there is no obvious evidence justifying such tuning of the migration scheme, it does \rev{mimic ``planet traps'' caused by} inverted gas pressure gradients \rev{to some degree.
Their existence is indicated by numerous observations of disk substructures~\citep[e.g.,][]{Andrews2018b}, which are suspected to be also common around very low-mass stars~\citep{Kurtovic2021,Pinilla2021}.}
This way out of the giant planet conundrum still requires relatively high initial solid disk masses $M_\mathrm{solid,0} \gtrsim \SI{66}{\mEarth} \approx \SI{2e-4}{\mSun}$.
The efficiency of \rev{planet core} formation could be further enhanced if large amounts of planetesimals are concentrated \rev{at} intermediate orbital distances, possibly at the water iceline~\citep{Drazkowska2017}.

Giant planet formation could be further facilitated if planet cores accrete not only planetesimals but also mm to cm-sized pebbles~\citep{Klahr2006,Ormel2010,Lambrechts2012,Voelkel2020}, but sufficiently high pebble fluxes are needed in order for growth to outweigh rapid inward migration~\citep{Bitsch2019a}.

We conclude that common theories of planet formation via core accretion do not predict planets like the gas giants around late M stars described here.
\rev{Giant planet growth via direct collapse in a gravitationally instable disk remains an alternative scenario to form these planets, provided that the mass infall rate onto the disk at any one time exceeded its turbulence-driven accretion rate~\citep{Boss1997}.
However, planets formed via disk instability are thought to form at large orbital radii and to be massive with $M\sim 10\,\mathrm{M_{Jup}}$~\citep[e.g.,][]{Adams1992,Kratter2010}.
It is thus difficult to reconcile the planets considered here with this scenario.
}

A larger observational sample and systematic parameter studies of various formation models will be required to reveal the physical mechanisms responsible for this enigmatic subpopulation of planets.
Stronger constraints can be expected from future microlensing surveys, which \rev{predominantly probe low-mass stars}~\citep{Gould2010,Zang2021,Hwang2021}.
The high angular resolutions achieved by extremely large telescopes will allow one to resolve and measure masses of the hosts of essentially all microlensing planets discovered to date.
Furthermore, astrometric data from Gaia Data Releases~3 and~4 will enable much better statistics on a sample of several thousand nearby M~dwarfs~\citep{Sozzetti2014,Perryman2014}.
For these stars, giant planets on short and intermediate orbits can then be characterized, which will greatly enhance the sample size for precise occurrence rate studies of giant planets around low-mass stars.

\subsection{A stellar-mass dependent drop in the period distribution}\label{sec:discussion_periods}
The present-day picture of planet-disk interactions generally leads to migration of planets toward the star.
Therefore, if and where they are stopped can in principle be constrained by the orbital period distribution of exoplanets.
Our sample provides new constraints to this topic, \rev{which was previously predominantly explored with transit surveys.
   \citet{Mulders2015a} found a stellar mass dependent drop in occurrence rates of inner planets from the \emph{Kepler} mission and determined a semi-major axis break point that scales as $\propto M_{\star}^{1/3}$.
The innermost planet of a given system appears to be located at a preferred orbital period~$\simeq$\SI{10}{\day}~\citep{Mulders2018a}.
As drivers of such a drop, migration traps~\citep{Plavchan2013} or the removal of inner planets are being discussed.
}

\revii{It seems} \rev{possible to utilize trends with stellar host mass} to pinpoint which process determines the location of the innermost planets.
\rev{\citet{Mulders2015a} favored planetary tides or the stellar corotation radius to match their $\propto M_{\star}^{1/3}$ scaling.
They further considered a} removal of planets by stellar tides \rev{or} dust sublimation in passive or viscously heated disks.
Here, we briefly discuss new insights gained on the possible origins of this scaling in light of our sample.

From Fig.~\ref{fig:period_hist_Mstar} it becomes apparent that the Bern model fails to correctly predict the stopping mechanism: synthetic planets commonly migrate closer to the star than their observed counterparts. The model includes an inner edge set at corotation radii based on observed \rev{stellar} rotation rates (\citealp{Venuti2017}, see the discussion in \citealp{Burn2021})\rev{.
The drop of the gas surface density at this location causes a \mbox{type-I} outward migration zone~\citep{Masset2006}, which \revii{should} foster a local pile-up of planets~\citep{Mulders2019}.
However, we observe only little change of the innermost periods in simulations with increased inner disk edge position~(see Appendix~\ref{sec:innerDiskEdgeTest}).}
\revii{As indicated in \citet{Schlecker2021b}, }\rev{we find that the presence of planets within the corotation radius} can be attributed to N-body interactions with planets further out:
Often, an inner planet gets locked in a mean-motion resonance with an outer migrating planet that pushes it further in.
This model outcome is in agreement with the result of \citet{Ataiee2021}, who use 2D hydrodynamic models instead of migration rate prescriptions to address this issue.
They found that commonly an inner edge is not able to stop migration of resonant chains of planets.
\revii{This is in agreement with the resonant chain-shaped period distribution of planets around solar-mass stars in \citet{Carrera2019}.}
\rev{Thus, the inner edge of the disk is only an efficient migration trap for solitary planets.
With our multiseed setup, such a scenario is rare.
}

A more efficient migration trap could be the inner edge of a \rev{nonionized} dead zone \citep{Gammie1996}.
Due to a change in ionization rate, the viscosity is expected to increase, which in turn leads to a lowered surface density and a pressure bump in a steady-state disk.
The typical temperature at which this happens is $\sim$\SI{1000}{\kelvin}~\citep{Flock2016,Flock2017}.
More detailed models were presented by \citet{Mohanty2018} and \citet{Jankovic2021}: They find pressure bumps outside the dust sublimation front.
Therefore, these pressure bumps can cause the first trap that resonant convoys of migrating planets would encounter.
According to \citet{Ataiee2021}, the \rev{dead zone} inner edge is efficient in halting migration of planets in resonant chains.
The scaling of the orbital period of the pressure bump found by \citet{Mohanty2018} is $\propto M_{\star}^{3q/4}$, where $q$ is the exponent with which the stellar accretion rate depends on stellar mass $q\sim 1.8$ \citep{Alcala2017}.
Such a steep dependency of a trap mechanism on stellar mass is required to explain the tentative evidence from our sample shown in the right panel of Fig.~\ref{fig:period_hist_Mstar}.
While the scaling found by \citet{Mohanty2018} looks promising to explain the trend, the absolute values of the found pressure bump locations are too large with orbital periods of $\sim$\SI{60}{\day} for solar-mass stars.
However, the work of \citet{Mohanty2018} does not include dust evolution, which would change the outcome.

\rev{A related effect that is included in the Bern model is} radial migration due to stellar tides \citep[following e.g.,][]{Benitez-Llambay2011}.
Similar to the consideration of \citet{Mulders2015}, they lead to the removal of the planets closest to their host stars.
\rev{In the simulations, tidal effects are at least partially responsible for shaping the period distribution at close orbits.}
Despite that, the observed distribution is not reproduced, which indicates that either tides are stronger in reality or that \rev{we are missing} another important mechanism.

The modeled stellar \rev{tides in \citet{Burn2021} include} the dependency on stellar mass and radius \citep[determined using the stellar evolution tracks by][]{Baraffe2015}.
\rev{Currently missing is the dependency on the ratio between the quadrupolar hydrostatic Love number and the} tidal dissipation quality factor $Q_{\star}$ \citep{Gallet2017}.
These parameters depend on the interior structure and dynamics of the star, which \rev{vary} with stellar mass.
A detailed analysis taking this into account was done by \citet{Strugarek2017}\rev{,} who find that tides can become efficient for solar-mass stars \rev{and} orbital periods below \SI{20}{\day}.
However, the dynamical tide and therefore a large portion of the effect is suppressed for fully convective M~dwarfs.
Therefore, fewer planets would be removed at \rev{small orbital periods.
This could explain the different distribution of planetary orbital periods for stars with $M_{\star}<\SI{0.4}{M_{\odot}}$.}

Another effect leading to the same outcome as tides are magnetic planet-star interactions. \citet{Strugarek2017} find that they can dominate over tides in some regimes, but the order of magnitude is comparable to tides.
For low stellar masses, magnetic effects are stronger than tides but are not efficient enough to lead to migration of planets on orbits with periods larger than a day.

Overall, our M~dwarf sample provides tentative evidence for a steep scaling of a migration trap with stellar mass that could be caused by a magnetically induced pressure bump.
Alternatively, efficient tidal migration could remove the innermost planets around larger stars but cease to be efficient for fully convective small stars.
These findings should be seen as further motivation for a model revision of the planetary orbital migration and trapping mechanisms already mentioned in \citet{Emsenhuber2021} and \citet{Schlecker2021}.

\subsection{Caveats and limitations}\label{sec:caveats}
Radial velocity surveys are not free from biases, and the CARMENES and HARPS surveys are no exception.
At the same time, no theoretical model can fully reflect all physical mechanisms relevant for planet formation, and simplifications have to be made.
In the following, we list a number of limitations of our study that could affect our conclusions.

\subsubsection{Selection effects}
    \rev{Spectroscopic binaries have been removed from the HARPS and CARMENES samples, respectively. \citet{Moe2021} showed that this may lead to overestimated giant planet occurrence rates for G-type stars.
    \revii{Reported close binary fractions are relatively constant across the M~dwarf mass range~\citep{Offner2022}. Thus, the removal of binary stars from the samples can} not lead to the apparent overabundance of giant planets around late M~dwarfs.
    Since all our simulated planet hosts are single stars, this selection effect does not impair our comparison of observed and synthetic populations.}

    One potential bias is the intensified observation of targets that already show a tentative signal.
    Such a signal shows earlier for more massive planets, and these planets are thus more likely to be eventually detected.
    The CARM$_{70}$ sample was selected out of a larger sample with about 340 targets based on the number of observations its stars received.
    Thus, ``human intervention bias'' may lead to an overprediction of more massive observed planets up to a factor of five, assuming there are no giant planets left in the rest of the sample \citep[see also][]{Sabotta2021}.
    \rev{Since we have no indication of an enhancement of this effect for the least massive stars, it should not affect our conclusions.}
    The analysis of the full CARMENES sample will result in more robust occurrence rates for giant planets.
\subsubsection{Single-planet approximation during biasing}
The \textit{NGM} population consists of multiplanet systems where most systems maintain several planets each.
In practice, their detectability not only depends on the individual planet properties but also on the combination of planets that occupy a system: The measured RV signal of a multiplanet system is a combination of contributions from the individual planets, and a successful disentanglement of these signals depends on their shape.
Nevertheless, when applying the detection bias to synthetic systems, we treat each planet as isolated and assign its detectability based on the survey sensitivity at its orbital period and minimum mass.
Injection-and-retrieval tests of each synthetic system might lead to a more realistic bias but would be computationally expensive and are not expected to have a significant impact on the statistical result.
\subsubsection{Simplifications of the formation model}
	Due to computational limits, there remain relevant model parameters, such as the viscous $\alpha$ parameter or the planetesimal size, that are not included in our parameter search.
    While the chosen values result in a good fit for the solar mass case, there might be additional trends with stellar mass that need to be explored in the future.
	In addition, our model does not yet include the evolution of solids in the disk and the consistent formation of planetesimals and seeds~\citep{Burn2021}.
    We note that the popular theoretical approximation of two grain sizes presented in \citet{Birnstiel2012} was derived for solar-type stars and its application to an order of magnitude lower stellar masses is not straightforward.

\section{Conclusions}\label{sec:conclusions}
We have compared synthetic planet populations computed with a core accretion formation model with a sample of planets around low-mass stars discovered by the HARPS and CARMENES RV surveys.
To correct for completeness, we performed injection tests on the actual RV time series of the surveys\rev{.
Instead of extrapolating beyond the observed planet sample, we biased the population of simulated systems according to the detection limits of the observed sample.}
We then statistically compared the actual and synthetic surveys in $\msini$-$P$-$M_\star$ space.
Our main findings are:
\begin{itemize}
    \item Theory and observations are in agreement for short-period rocky planets, which form the largest population. Their observed detection rates, planetary mass function, \rev{and orbital period distribution} are consistent with our simulations.
    \item Observed detections of giant planets around late (\SI{<0.5}{\Msun}) M~dwarfs \rev{might be }indicative of a break in the giant planet occurrence as a function of host star mass. The existence of these planets \rev{cannot} be reconciled with our model, although the discrepancy is reduced when planet migration is suppressed.
    \item The observed and synthetic planetary mass functions diverge for intermediate masses (\SIrange{30}{200}{\mEarth}). More sophisticated treatments of planetary gas accretion, in particular \rev{those} that take into account 3D effects, are possibly needed for convergence.
	\item The observed \rev{orbital period distribution depends} on the stellar mass\rev{ with a paucity of very short-period planets around stars $\gtrsim \SI{0.4}{M_{\odot}}$}.
    While the model reproduces the distribution for \rev{less massive stars, it fails} to remove planets or halt planetary migration efficiently enough for \rev{earlier M~dwarfs}.
    Candidate mechanisms to produce the observed trend are planet trapping due to a \rev{dead zone} inner edge or stellar tides.
\end{itemize}
Both the \rev{inability to explain the} existence of giant planets around very low-mass stars and the difference in the period distributions \rev{suggest that state-of-the-art planet formation models are still missing a complete picture of} planet migration, or rather of the local disk conditions to which it is highly sensitive.
\rev{Constraints on the abundance of substructures in disks around very low-mass stars through high angular resolution observations will be particularly insightful here.}

With \Nstars\ stars and \Nplanets\ planets, the statistical power of our sample is still limited.
It will improve upon the completion of the CARMENES GTO survey, and beyond that by including the imminent results from astrometry and microlensing campaigns.
On the theoretical side, future model improvements will allow one to study the physical mechanisms responsible for the discrepancies presented in this paper.

Our findings underscore the different conditions in protoplanetary disks around different stellar spectral types, which has a measurable impact on the outcome of planet formation.
M~dwarfs are not just small Suns.

\begin{acknowledgements}
We wish to thank Pedro Amado, \rev{Hans Baehr,} Jose Caballero, Mario Flock, Andrew Gould, Diana Kossakowski, \rev{Nicolas Kurtovic,} \revii{Kaitlin Kratter,} Rafael Luque, \rev{Ilaria Pascucci,} Ansgar Reiners, and Andreas Quirrenbach for stimulating discussions.
We also thank the anonymous referee for their insightful comments that improved the manuscript.
This work was supported by the Deutsche Forschungsgemeinschaft (DFG) \rev{via} the DFG Research Unit FOR2544 “Blue Planets around Red Stars” (RE~2694/8-1)\rev{, as well as through the DFG priority programs SPP~1992: “Exoplanet Diversity” (KL~1469/17-1) and SPP~1833 “Building a Habitable Earth" (KL~1469/13-(1-2)).}
\rev{This material is based upon work supported by the National Aeronautics and Space Administration
under Agreement No. 80NSSC21K0593 for the program ``Alien Earths.'' The results reported herein
benefitted from collaborations and/or information exchange within NASA’s Nexus for Exoplanet System
Science (NExSS) research coordination network sponsored by NASA’s Science Mission Directorate.}
This research has made use of the SIMBAD database, operated at CDS, Strasbourg, France.
This research has made use of the VizieR catalogue access tool, CDS, Strasbourg, France (DOI:10.26093/cds/vizier). The original description of the VizieR service was published in 2000, A\&AS 143, 23.
\end{acknowledgements}


\bibliographystyle{aa} 
\bibliography{PhD,addLit} 

\begin{appendix} 

\section{Planet and stellar sample}

\begin{table*}
\caption[]{Observed stars used in this study. Adapted from~\citet{Sabotta2021} and \citet{Bonfils2013}.}
\label{table:starlist}
\centering     \onehalfspacing \tiny
\begin{tabular}{l c l c l c l c}
\hline\hline
\noalign{\smallskip}
Name & Mass (M$_\odot$)  & Name & Mass (M$_\odot$)  &  Name & Mass (M$_\odot$)  &Name & Mass (M$_\odot$) \\
\noalign{\smallskip}
\hline
\noalign{\smallskip}
\object{BD-055715}$^a$ & 0.46 & \object{Gl402} & 0.26 & \object{HIP31292} & 0.31 & \object{KarmnJ17378+185}$^a$ & 0.42 \\
\object{BD-073856} & 0.61 & \object{Gl413.1} & 0.46 & \object{HIP31293} & 0.43 & \object{KarmnJ17578+046}$^a$ & 0.17 \\
\object{BD+053409}$^a$ & 0.51 & \object{Gl433} & 0.47 & \object{KarmnJ00051+457} & 0.50 & \object{KarmnJ18174+483} & 0.58 \\
\object{BD+092636} & 0.53 & \object{Gl438} & 0.33 & \object{KarmnJ00067-075}$^a$ & 0.11 & \object{KarmnJ19169+051N}$^a$ & 0.45 \\
\object{BD+63869} & 0.55 & \object{Gl447} & 0.17 & \object{KarmnJ00183+440} & 0.34 & \object{KarmnJ19346+045} & 0.55 \\
\object{CD-44863} & 0.22 & \object{Gl465} & 0.34 & \object{KarmnJ01025+716} & 0.47 & \object{KarmnJ20305+654} & 0.36 \\
\object{G108-21} & 0.23 & \object{Gl479} & 0.43 & \object{KarmnJ01026+623} & 0.50 & \object{KarmnJ21164+025} & 0.35 \\
\object{G192-013} & 0.25 & \object{Gl480.1} & 0.18 & \object{KarmnJ01125-169}$^a$ & 0.15 & \object{KarmnJ21348+515} & 0.46 \\
\object{G264-018A} & 0.53 & \object{Gl514} & 0.53 & \object{KarmnJ02222+478} & 0.55 & \object{KarmnJ21466+668} & 0.25 \\
\object{GJ1061} & 0.12 & \object{Gl526} & 0.5 & \object{KarmnJ02362+068}$^a$ & 0.25 & \object{KarmnJ22021+014}$^a$ & 0.57 \\
\object{GJ1065} & 0.19 & \object{Gl536} & 0.52 & \object{KarmnJ02442+255} & 0.34 & \object{KarmnJ22114+409} & 0.15 \\
\object{GJ1068} & 0.13 & \object{Gl551} & 0.12 & \object{KarmnJ02530+168} & 0.09 & \object{KarmnJ22115+184} & 0.54 \\
\object{GJ1123} & 0.21 & \object{Gl555} & 0.28 & \object{KarmnJ03133+047}$^a$ & 0.16 & \object{KarmnJ22252+594} & 0.34 \\
\object{GJ1125} & 0.29 & \object{Gl569} & 0.49 & \object{KarmnJ03463+262} & 0.57 & \object{KarmnJ22532-142}$^a$ & 0.32 \\
\object{GJ1224} & 0.14 & \object{Gl581} & 0.3 & \object{KarmnJ04153-076}$^a$ & 0.27 & \object{KarmnJ23216+172}$^a$ & 0.39 \\
\object{GJ1232} & 0.20 & \object{Gl588} & 0.47 & \object{KarmnJ04290+219} & 0.64 & \object{KarmnJ23351-023} & 0.13 \\
\object{GJ1236} & 0.22 & \object{Gl618.1} & 0.39 & \object{KarmnJ04376+528} & 0.54 & \object{KarmnJ23381-162} & 0.35 \\
\object{GJ1256} & 0.19 & \object{Gl643} & 0.21 & \object{KarmnJ04588+498} & 0.58 & \object{KarmnJ23419+441} & 0.15 \\
\object{GJ2066} & 0.46 & \object{Gl667} & 0.3 & \object{KarmnJ06103+821} & 0.40 & \object{LHS1134} & 0.2 \\
\object{Gl1} & 0.39 & \object{Gl674} & 0.35 & \object{KarmnJ06105-218}$^a$ & 0.52 & \object{LHS1481} & 0.17 \\
\object{Gl12} & 0.22 & \object{Gl680} & 0.47 & \object{KarmnJ06548+332} & 0.34 & \object{LHS1513} & 0.09 \\
\object{Gl145} & 0.32 & \object{Gl682} & 0.27 & \object{KarmnJ08413+594} & 0.12 & \object{LHS1723} & 0.17 \\
\object{Gl176} & 0.50 & \object{Gl693} & 0.26 & \object{KarmnJ09143+526} & 0.59 & \object{LHS1731} & 0.27 \\
\object{Gl191} & 0.27 & \object{Gl729} & 0.17 & \object{KarmnJ09144+526} & 0.59 & \object{LHS1935} & 0.29 \\
\object{Gl203} & 0.19 & \object{Gl754} & 0.18 & \object{KarmnJ10122-037}$^a$ & 0.53 & \object{LHS288} & 0.1 \\
\object{Gl205}$^b$ & 0.60 & \object{Gl803} & 0.75 & \object{KarmnJ10289+008}$^a$ & 0.41 & \object{LHS337} & 0.15 \\
\object{Gl213} & 0.22 & \object{Gl832} & 0.45 & \object{KarmnJ10482-113} & 0.12 & \object{LHS3583} & 0.4 \\
\object{Gl250} & 0.45 & \object{Gl87} & 0.45 & \object{KarmnJ10564+070}$^a$ & 0.15 & \object{LHS3746} & 0.24 \\
\object{Gl273} & 0.29 & \object{Gl877} & 0.43 & \object{KarmnJ11033+359} & 0.34 & \object{LP771-95} & 0.24 \\
\object{Gl285} & 0.31 & \object{Gl880} & 0.58 & \object{KarmnJ11054+435} & 0.35 & \object{LP816-60} & 0.23 \\
\object{Gl299} & 0.14 & \object{Gl887} & 0.47 & \object{KarmnJ11417+427} & 0.35 & \object{LP819-052}$^a$ & 0.18 \\
\object{Gl300} & 0.26 & \object{Gl908} & 0.42 & \object{KarmnJ11421+267} & 0.41 & \object{LTT9759} & 0.54 \\
\object{Gl341} & 0.55 & \object{HD165222}$^a$ & 0.44 & \object{KarmnJ11511+352} & 0.45 & \object{NLTT56083} & 0.29 \\
\object{Gl357} & 0.33 & \object{HD168442} & 0.59 & \object{KarmnJ12123+544S} & 0.57 & \object{Ross1020} & 0.26 \\
\object{Gl358} & 0.42 & \object{HD199305} & 0.51 & \object{KarmnJ14257+236W} & 0.60 & \object{Ross104} & 0.36 \\
\object{Gl367} & 0.49 & \object{HD260655} & 0.45 & \object{KarmnJ16167+672S} & 0.60 & \object{Wolf437}$^a$ & 0.31 \\
\object{Gl388} & 0.42 & \object{HD97101B} & 0.54 & \object{KarmnJ16303-126}$^a$ & 0.29 & ~ & ~\\
\noalign{\smallskip}
\hline
\end{tabular}
\singlespacing
    (a) Duplicate star removed from the HARPS sample \\
    (b) Duplicate star removed from the CARMENES sample
\end{table*}

\begin{table*}
\caption[]{Observed planets used in this study. Adapted from~\citet{Sabotta2021} and \citet{Bonfils2013}.}
\label{table:planets1mchapter}
\centering     \onehalfspacing \tiny
\begin{tabular}{l l l c c l l }
\hline\hline
\noalign{\smallskip}
Simbad & Karmn.  & ~ & $P$ & $\msini$ & Ref.  & Ref.  \\
Identifier & ID & ~ & (d) & (\si{M_\oplus})& Discovery & Param.  \\
\noalign{\smallskip}
\hline
\noalign{\smallskip}
\object{YZ Cet }           & J01125-169  & c &   3.060   &  $1.14^{+0.11}_{-0.10}$  & Ast17   &  Sto20a \\
\object{YZ Cet  }          & J01125-169  & d &   4.656   &  $1.09^{+0.12}_{-0.12}$  &  Ast17  & Sto20a  \\
\object{Teegarden's Star } & J02530+168  & b &   4.910   &  $1.05^{+0.13}_{-0.12}$  &  Zec19  &  Zec19  \\
\object{Teegarden's Star } & J02530+168  & c &   11.41   &  $1.11^{+0.16}_{-0.15}$  &  Zec19  &  Zec19  \\
\object{CD Cet   }         & J03133+047  & b &   2.291   &  $3.95^{+0.42}_{-0.43}$  &  Bau20  &  Bau20  \\
\object{HD 285968 }        &  J04429+189 & b &   8.78     & $9.06^{+1.54}_{-0.7}$   & For09 & Tri18 \\
\object{HD 265866  }       & J06548+332  & b &   14.24   & $4.00^{+0.40}_{-0.40}$   &  Sto20b &  Sto20b \\
\object{G 234-45    }      & J08413+594  & b &   203.6  &  $147^{+7.0}_{-7.0}$      &  Mor19 &  Mor19 \\
\object{HD 79211     }     & J09144+526  & b &   24.45    &  $10.3^{+1.5}_{-1.4}$   &  Gon20 &  Gon20 \\
\object{HD 95735    }      & J11033+359  & b &   12.95   &  $2.69^{+0.25}_{-0.25}$  &  Dia19 &  Sto20b \\
\object{Ross 1003}         & J11417+427  & b &   41.38   &  $96.6^{+1.3}_{-1.0}$    &  Hag10&  Tri20 \\
\object{Ross 1003}         & J11417+427  & c &   532.02  &  $72.1^{+0.3}_{-7.0}$     &  Tri18 &  Tri20 \\
\object{Ross 905}          & J11421+267  & b &   2.644    &  $21.4^{+0.20}_{-0.21}$ &  But04 &  Tri18 \\
\object{HD 238090}         & J12123+544S & b &   13.67   &  $6.89^{+0.92}_{-0.95}$  &  Sto20b &  Sto20b \\
\object{Wolf 437}$^a$      & J12479+097  & b &   1.467    &  $2.82^{+0.11}_{-0.12}$ &  Tri21 &  Tri21\\
\object{Ross 1020}         & J13229+244  & b &   3.023    &  $8.0^{+0.5}_{-0.5}$    &  Luq18 &  Luq18 \\
\object{BD-07 4003}        & J15194-077  & b &   5.37    & $15.20^{+0.22}_{-0.27}$ & Bon05 & Tri18 \\
\object{BD-07 4003}        & J15194-077  & c &   12.92    & $5.65^{+0.39}_{-0.24}$ & Udr07 & Tri18 \\
\object{BD-07 4003}        & J15194-077  & e &   3.15     & $1.66^{+0.24}_{-0.16}$ & May09 & Tri18 \\
\object{HD 147379}         & J16167+672S & b &   86.54    &  $24.7^{+1.8}_{-2.4}$   &  Rei18&  Rei18 \\
\object{BD-12 4523}        & J16303-126  & b &   1.26    &  $1.92^{+0.37}_{-0.37}$  &  Wri16 & Sab21 \\
\object{BD-12 4523}        & J16303-126  & c &   17.87   &  $4.15^{+0.37}_{-0.37}$  &  Wri16 & Wri16 \\
\object{BD+18 3421}        & J17378+185  & b &   15.53   &  $6.24^{+0.58}_{-0.59}$  &  Lal19&  Lal19\\
\object{HD 180617}         & J19169+051N & b &   105.9  &  $12.2^{+1.0}_{-1.4}$     &  Kam18&  Kam18\\
\object{LSPM J2116+0234}   & J21164+025  & b &   14.44   &  $13.3^{+1.0}_{-1.1}$    &  Lal19 &  Lal19\\
\object{G 264-12}          & J21466+668  & b &   2.305   &  $2.50^{+0.29}_{-0.30}$  &  Ama21&  Ama21\\
\object{G 264-12}          & J21466+668  & c &   8.052    &  $3.75^{+0.48}_{-0.47}$ &  Ama21&  Ama21\\
\object{L 788-37}          & J22137-176  & b &   3.651    &  $7.4^{+0.5}_{-0.5}$    &  Luq18&  Luq18 \\
\object{G 232-70}          & J22252+594  & b &   13.35   &  $16.6^{+0.94}_{-0.95}$   &  Nag19&  Nag19\\
\object{BD-15 6290}        & J22532-142  & b &   61.08   &  $761^{+1.0}_{-1.0}$     & Del98 &  Tri18\\
\object{BD-15 6290}        & J22532-142  & c &   30.13   &  $242^{+0.7}_{-0.7}$     & Mar01 &  Tri18\\
\object{CD-31 9113}        &  ~          & b &   7.37    & 5.49 & Bon13 & Del13 \\
\object{CD-46 11540}       & ~           & b &   4.69     & 11.39 & Bon07 & Boi11 \\
\object{HD 156384C}        & ~           & b &   7.20     & $5.6^{+1.4}_{-1.3}$ & Bon13 & Ang13 \\
\object{HD 156384C}        & ~           & c &   28.14    & $3.8^{+1.5}_{-1.2}$ & Bon13 & Ang13 \\
\noalign{\smallskip}
\hline
\end{tabular}\singlespacing 
\tablebib{\tiny
    Ama21: \citealt{Amado2021};
    Ang13: \citealt{Anglada2013};
    Ast17: \citealt{AstudilloDefru2017};
    Bau20: \citealt{Bauer2020};
    Boi11: \citealt{Boisse2011};
    Bon05: \citealt{Bonfils2005};
    Bon07: \citealt{Bonfils2007};
    Bon13: \citealt{Bonfils2013};
    But04: \citealt{Butler2004};
    Del98: \citealt{Delfosse1998};
    Del13: \citealt{Delfosse2013};
    Dia19: \citealt{Diaz2019};
    For09: \citealt{Forveille2009};
    Gon20: \citealt{GonzalezAlvarez2020};
    Hag10: \citealt{Haghighipour2010};
    Kam18: \citealt{Kaminski2018};
    Lal19: \citealt{Lalitha2019};
    Luq18: \citealt{Luque2018};
    Mar01: \citealt{Marcy2001};
    May09: \citealt{Mayor2009};
    Mor19: \citealt{Morales2019};
    Nag19: \citealt{Nagel2019};
    Rei18: \citealt{Reiners2018};
    Riv05: \citealt{Rivera2005};
    Sto20a: \citealt{Stock2020};
    Sto20b: \citealt{Stock2020a};
    Tri18: \citealt{Trifonov2018};
    Tri20: \citealt{Trifonov2020b} ;
    Tri21: \citealt{Trifonov2021};
    Udr07: \citealt{Udry2007};
    Wri16: \citealt{Wright2016};
    Zec19: \citealt{Zechmeister2019}.\\
    \tablefoottext{a}{For the transiting planet GJ\,486b (J12479+097), its determined mass is listed instead of its minimum mass.}
}
\end{table*}

\rev{
\section{Influence of the inner disk edge on orbital periods}\label{sec:innerDiskEdgeTest}
In Sect.~\ref{sec:results_orbital_periods} we showed that our model produces synthetic detection rates that decrease approximately log-linearly with orbital period.
\revii{This shape is caused by the combination of the underlying synthetic period distribution and applying the detection bias to the synthetic sample, which favors short-period planets over more distant ones.
The relative occurrence rates differ} from the \emph{Kepler} sample, which peaks at \SI{\sim10}{\day} and declines for shorter orbital periods~\revii{\citep{Howard2012,Mulders2015a,Carrera2019}}.
We observed a similar departure from a linear trend at the shortest orbital periods in the HARPS\&CARM$_{70}$ sample for the more massive stars.

This raises the question if the continuous decline in the model is merely defined by a numerical inner disk edge that is too close to the star.
To explore this, we examine the period distribution of a comparison population with inner edges at periods $>\SI{7.3}{\day}$ instead of the nominal distribution with a median of \SI{4.74}{\day}~(Fig.~\ref{fig:periods_inner_edge_test}).
Differences between the populations are visible, but no clear deviation from the original trend appears.
In particular, the increased distance of the disk edge \revii{is not sufficient to overcome the effect of the detection bias and to }cause a break in the distribution with a drop at small periods.
An extended cavity hence fails to fully explain the observed lack of inner planets around the more massive stars in our sample.
\begin{figure*}
    \centering
    \includegraphics[width=\hsize]{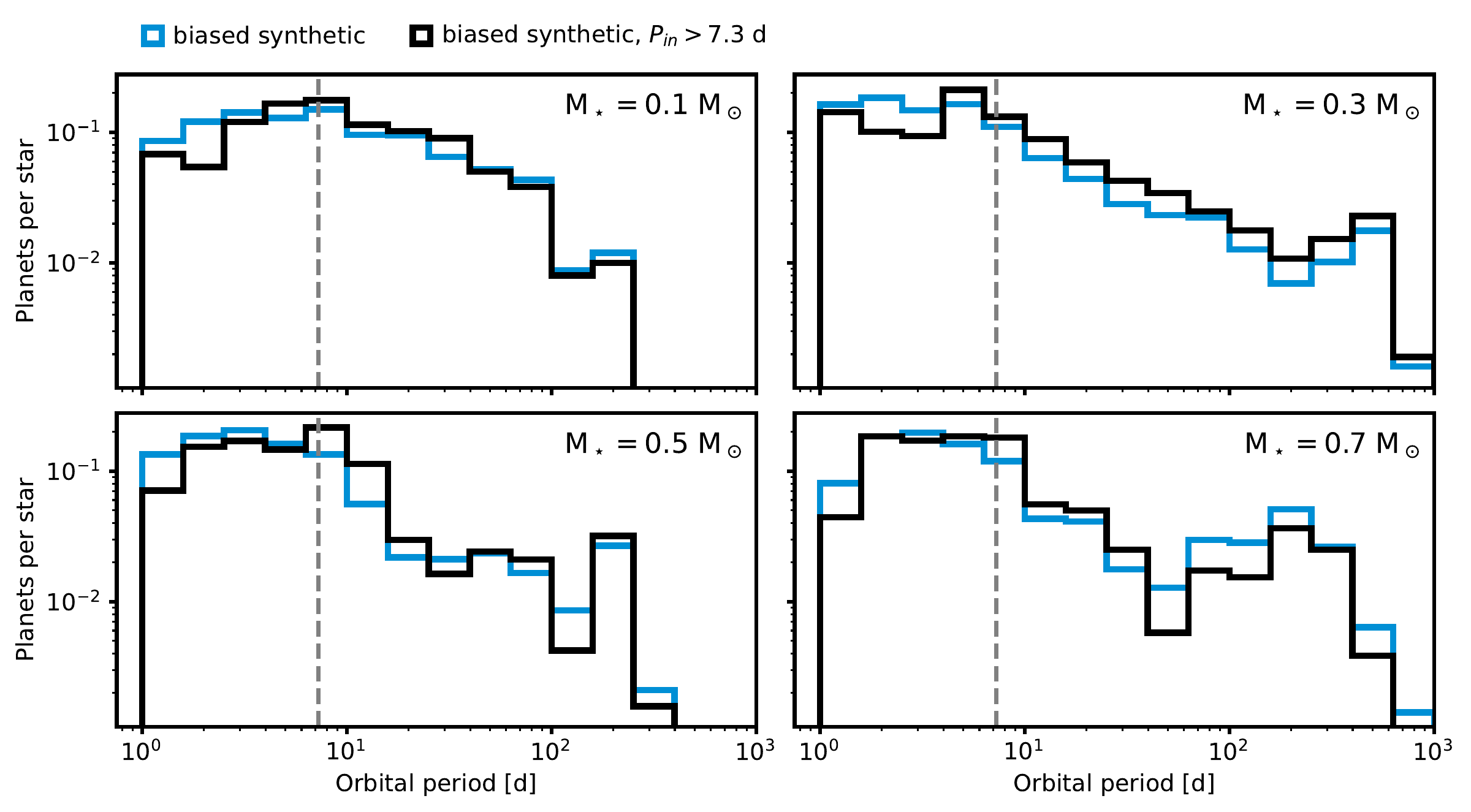}
    \caption[]{\rev{Dependence of the orbital period distribution on the position of the inner disk edge.
    For each stellar host mass bin, we show the nominal biased synthetic population and a comparison sample with inner disk edges greater than \SI{7.3}{\day} (dashed gray lines), which is the 68th percentile of the distribution. While differences occur, there is no clear deviation from the approximate linear trend.}}
    \label{fig:periods_inner_edge_test}
\end{figure*}
}
\end{appendix}
\end{document}